\documentclass[aps,prl,twocolumn,amsmath,amssymb,superscriptaddress,floatfix]{revtex4-2}

\usepackage{amsmath}
\usepackage{amssymb}
\usepackage{amsthm}
\usepackage{mathtools}
\usepackage{mathdots}
\usepackage{amsfonts}
\usepackage{bbm}
\usepackage{bm}
\usepackage{xr}
\usepackage{bbold}
\usepackage{epsfig,color,dsfont,upgreek,physics}
\usepackage{mathrsfs}
\usepackage{multirow}
\usepackage{graphicx}
\usepackage{dcolumn} 
\usepackage{float} 
\usepackage[driverfallback=dvipdfm]{hyperref} 
\usepackage{longtable}
\usepackage{ulem}   
\allowdisplaybreaks
\usepackage{outlines} 

\newcommand{\bP}{{\textbf{P}}}
\newcommand{\bs}[1]{{\boldsymbol{#1}}}

\newcommand{\s}{{\sigma}}

\newcommand{\sgn}{\mathcal{\text{sgn}}}

\newcommand{\eff}{\mathcal{\text{eff}}}
\newcommand{\orb}{\mathcal{\text{orb}}}
\newcommand{\band}{\mathcal{\text{band}}}

\renewcommand\[{\begin{equation}}
\renewcommand\]{\end{equation}} 

\begin{document}

\title{Upper critical in-plane magnetic field in quasi-2D layered superconductors}

\author{Huiyang Ma}
\affiliation
{National High Magnetic Field Laboratory, Tallahassee, Florida 32310, USA}
\affiliation{Department of Physics,
Florida State University, Tallahassee, Florida 32306, USA}

\author{Dmitry V. Chichinadze}
\affiliation
{National High Magnetic Field Laboratory, Tallahassee, Florida 32310, USA}
\affiliation{Department of Physics, Washington University in St. Louis, St. Louis, Missouri 63160, USA}

\author{Cyprian Lewandowski}
\affiliation
{National High Magnetic Field Laboratory, Tallahassee, Florida 32310, USA}
\affiliation{Department of Physics,
Florida State University, Tallahassee, Florida 32306, USA}
\begin{abstract}
The study of the interplay of applied external magnetic field and superconductivity has been invigorated by recent works on Bernal bilayer and rhombohedral multilayer graphene. These studies, with and without proximitized spin-orbit coupling, have opened up a new frontier in the exploration of unconventional superconductors as they offer a unique platform to investigate superconductivity with high degree of in-plane magnetic field resilience and even magnetic field-induced superconductivity. Here, we present a framework for analyzing the upper critical in-plane magnetic field data in multilayer superconductors.  
Our framework relies on an analytically tractable superconducting pairing model that captures the normal state phenomenology of these systems and applies it to calculate the relationship between the upper critical field $H_{c2}$ and the corresponding critical temperature $T_{c}$. We study the $H_{c2}-T_{c}$ critical curve as a function of experimental parameters (Ising and Rashba spin-orbit coupling) and depairing mechanisms (Zeeman and orbital coupling) for both spin-singlet and spin-triplet pairing. By applying our framework to analyze four recent Bernal bilayer graphene-WSe$_2$ experiments \cite{Zhang2023_Enhanced_SC_proximitized_BBG,Zhang2025-ty, Holleis2025_Nemacity_BBG,Tingxin_Li2024_BBG_WSe2}, we identify an apparent discrepancy between fitted and measured spin-orbit parameters, which we propose can be explained  by an enhancement of the Land\'e g factor in the Bernal bilayer graphene experiments.

\end{abstract}
\date{\today}
\maketitle

\section{Introduction}

\begin{figure}[t!]
    \centering
    \includegraphics[width=\linewidth]{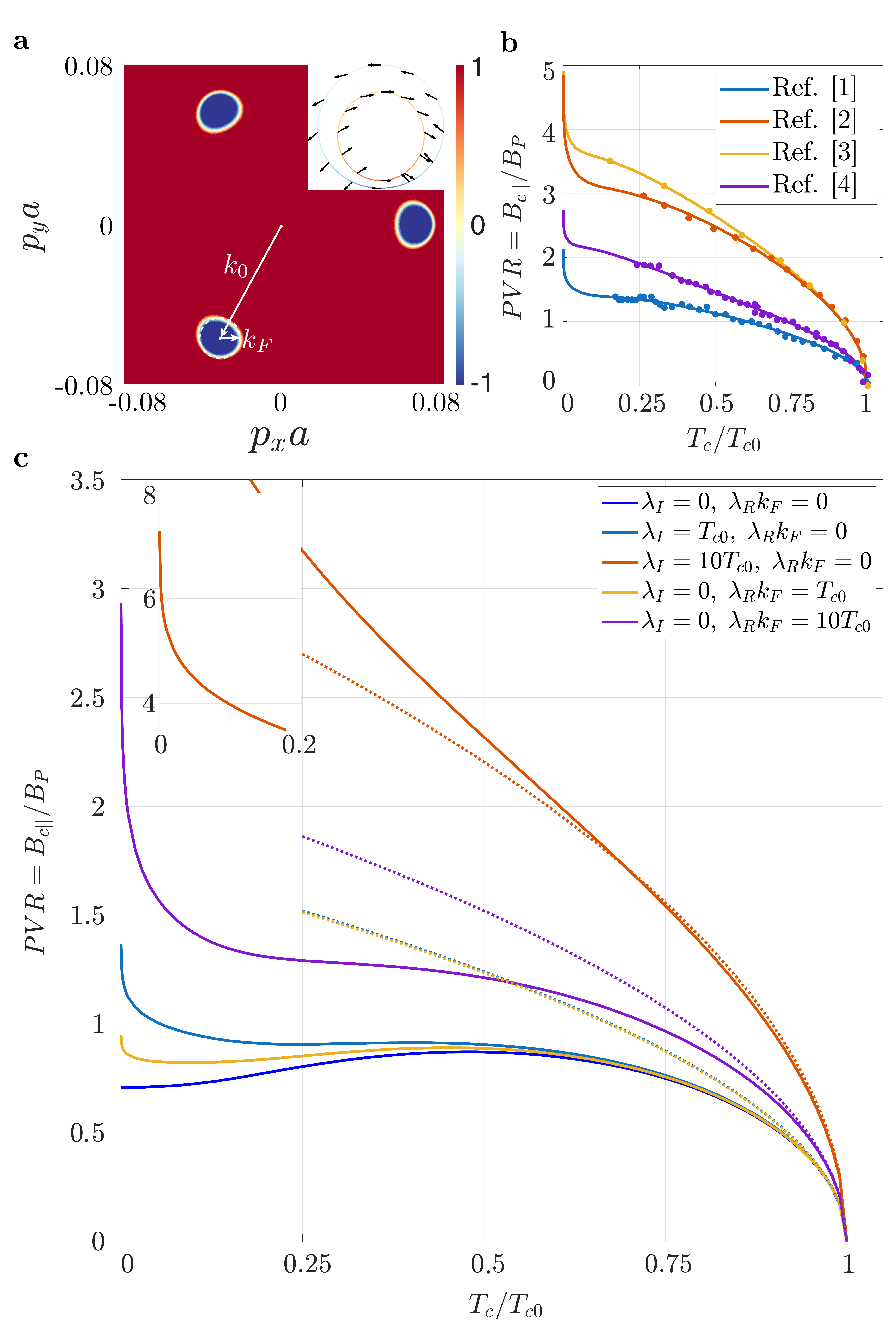}
    \caption{\textbf{a}: Equipotential lines of BBG shows three low-density pockets arising from trigonal warping together with approximate parametrization of the relevant pocket momenta (see text). The inset shows one such (idealized to a circular shape) pocket spin-split with expected spin textures (here the out-of-plane component) under the in-plane magnetic field and SOCs. \textbf{b}: The ratio (PVR) of the upper critical field to the Pauli limiting value (see text for definition). We fit the experimental data extracted from \textbf{i} \cite{Zhang2023_Enhanced_SC_proximitized_BBG} (blue), \textbf{ii} \cite{Zhang2025-ty} (red), \textbf{iii} \cite{Holleis2025_Nemacity_BBG} (yellow), and \textbf{iv} \cite{Tingxin_Li2024_BBG_WSe2} (purple). The fitted parameters are attached in Table \ref{tab:fitting result}.  \textbf{c}: Upper critical field for spin singlet 2D superconductor with Ising and Rashba SOCs for a range of parameters. Ising SOC shows an enhancement to $H_{c2}$, while the Rashba SOC are less relevant to the upper critical field except at the zero temperature limit.
    }
    \label{fig1 singlet}
\end{figure}

Recent years have been marked by a surge in interest in the effects of electron correlations in ultra-thin film systems with the discovery and synthesis of graphene \cite{graphene_Novoselov_2004,graphene_Kim_2005}, atomically thin 2D transition metal dichalcogenides (TMDs) \cite{Kin_Fai_MoS2_PRL_2010, Xiaodong_Di_Xiao_PRL2012}, and few-layer graphene heterostructures \cite{RotenbergScienceBBG2006}. These milestones led to a variety of important results and significantly advanced our understanding of non-superconducting ordered electronic states \cite{Huang2017CrI3, Cao2018Insulator, Park2023_FCI1, Cai2023_FCI1, Zeng2023_FCI, Tingxin_FCI_PRX_2023, Kang2024FractionalSpinHall, Redekop2024FCI_Imaging} as well as contributing to a list of new superconductors \cite{Cao2018SC, Xia2025,xu2025signaturesunconventionalsuperconductivitynear}. At the same time, there has been a revival of interest in a well-established area \cite{WHHPhysRev1966, FischerHelvetica1973, KlemmLutherBeasley, Fischer1978} investigating the interplay between superconductivity and the applied external magnetic field. In particular, recent discoveries of severe Pauli limit violation in bulk \cite{ShengScience2019, ShengRanNatPhys2019, AokiJPSJ2019, KnebelJPSJ2019, ThomasSciAdv2020, AjeeshPRX2023, TonyPNAS2024, Wu2024threedimensionalquantumcriticality}, thin film \cite{Dung_Sci_Adv_2023, Sun_Nickelates_Adv_Mat_2023, Dung2025, Badih2025} and 2D \cite{Zhou2021, AndreaBBGSCScience2022, Zhang2023_Enhanced_SC_proximitized_BBG, Holleis2025_Nemacity_BBG} superconductors attracted much attention because of links to possible mechanisms of field-reinforced or field-induced superconductivity \cite{JaccarinoPeterPRL1962, FischerHelvetica1973, GilAndyMacScience2001, BelitzKirkpatrickPRB2003, MonthouxGilNature2007, YellandNatPhys2011, BrandoMalteRMP2016}, triplet superconducting order parameters \cite{ShengScience2019, Cao2021, AndreaBBGSCScience2022} with the promise of technological applications, and a possibility of identifying superconducting order parameters by their upper critical field dependence on temperature.

The discovery of superconductivity in 2D graphene-based heterostructures has combined these two areas of research. While significant understanding of the normal (parent) states of these superconductors has been developed, the intrinsic properties of those superconductors, e.g., the superconducting gap symmetries or pairing mechanism, remain largely undetermined. For some graphene-based heterostructures, superconductivity seems to appear in the absence of spin-orbit coupling (SOC) in the system \cite{Cao2018SC, Zhou2021, AndreaBBGSCScience2022}, while for others, the presence of SOC is necessary to stimulate superconductivity \cite{Zhang2023_Enhanced_SC_proximitized_BBG, Holleis2025_Nemacity_BBG, Zhang2025-ty, Yang2025-SC4}. A complex interplay of SOC, superconducting gap symmetry, pairing mechanism, and the external magnetic field makes the interpretation of the upper critical measurements in such materials challenging, as in the case of superconductors with broken inversion symmetry \cite{AgterbergPRL2004, JinPRB2025}. 

Building upon these early works of Refs. \cite{EdelsteinPRL1995, GorkovRashbaPRL2001, GorkovBarzykin2002, AgterbergPRL2004, SergienkoPRB2004, AgterbergVorticesPRL2005, DimitrovaPRB2007, KTLawIsingScience2015, Smidman_2017, FuldeJPSJ2017, Ilic2017, ShafferPRB2020, JinPRB2025, WangPRL2019-TypeIIIsing}, here we partially resolve this challenge by providing a framework that, through consideration of in-plane magnetic field response of superconductivity, helps constrain the expected pairing symmetries of the system. To make our results general and analytically tractable, we develop an effective model designed to capture the phenomenology of superconductivity in a large class of recent 2D van der Waals (vdW) materials. Specifically, we consider near-circular Fermi surfaces, as seen in Bernal bilayer graphene (BBG) at low carrier density that are paired with their corresponding inversion symmetric partners, in the presence of substrate-induced Rashba and Ising SOCs, orbital and Zeeman depairing mechanisms - Fig. ~\ref{fig1 singlet}a. We derive our effective microscopic model starting from a faithful low-energy 4-band (8-band in the presence of SOC) model for BBG. However, we emphasize that our derivation scheme and low-energy model can also be applied to other 2D superconductors, such as rhombohedral multilayer graphene and TMDs.  We employ this model to calculate the upper critical field dependence on temperature, $H_{c2}(T)$, and find distinctly different types of behavior for singlet and triplet order parameters. We discuss the limitations of our approach and apply our methodology to the $H_{c2}(T)$ data in BBG from multiple experiments \cite{Zhang2023_Enhanced_SC_proximitized_BBG,Zhang2025-ty, Holleis2025_Nemacity_BBG,Tingxin_Li2024_BBG_WSe2}, see Fig.~\ref{fig1 singlet}b. There, we find that SOC is dominated by the Ising contribution, in accordance with experimental interpretation \cite{AndreaBBGSCScience2022, Holleis2025_Nemacity_BBG}. Unexpectedly, we also find that the experimental trends in BBG heterostructures with TMD-induced SOC can be interpreted as arising from a system with an enhanced $ g$-factor (larger than 2) -- a phenomenon recently discussed also in the context of Hofstadter physics in twisted TMDs \cite{XiaoyuOskar_TMD_PRB_2024}.

\section{The effective low-energy model}

To study the effects of SOCs, orbital and Zeeman depairing on the upper critical field of a 2D superconductor, we consider an effective low-energy model, motivated by prior theoretical studies in Ref. \cite{EdelsteinPRL1995, GorkovRashbaPRL2001, GorkovBarzykin2002, AgterbergPRL2004, SergienkoPRB2004, AgterbergVorticesPRL2005, DimitrovaPRB2007, KTLawIsingScience2015, Smidman_2017, FuldeJPSJ2017, Ilic2017, ShafferPRB2020, JinPRB2025, WangPRL2019-TypeIIIsing}, which is described by a quadratic Hamiltonian $H = \sum_{\bs{k},\xi} \Psi^{\dagger}_{\bs{k},s} h_{\xi} \left(\bs{k}\right) \Psi_{\bs{k},s'}$, where 
\begin{equation}
    h_{\xi} \left(\bs{k}\right) = \left( \epsilon_{\xi}\left(\bs{k}\right)-z_{0} \right)s_{0} + \left( \bs{g}_{I}+\bs{g}_{R}+\bs{b} \right) \cdot \bs{s}.
    \label{reduced_free_particle_h}
\end{equation}
Here $\epsilon_{\xi}\left(\bs{k}\right)$ is the quasiparticle dispersion in the valley $\xi$ in the absence of magnetic field and SOC, $\bs{g}_I =\frac{1}{2}\xi\lambda_I\hat{z}$, $\bs{g}_R = \frac{1}{2}\lambda_R \left(k_y, -k_x, 0\right)$, $\bs{b}=\frac{1}{2}g\mu_BB\hat{x}$ -- is the external in-plane magnetic field, $\hat{x},\;\hat{z}$ -- are the unit vectors in the $x-$ and $z-$direction. The term   $z_0=g_{\orb}\mu_BB$  represents the effective kinetic energy from inter-layer coupling generated by in-plane magnetic field, and $\xi = \pm 1$ stands for the valley index. Here $\lambda_{I}$ and $\lambda_R$ define magnitudes of the substrate-induced Ising and Rashba SOCs, respectively. Note that magnitudes of $\lambda_{I}$,  $\lambda_R$, $g$, and $g_{orb}$ can also  depend on the strength of the external displacement field, as we show later in the text and in SI due to the process of projecting the full Hamiltonian onto the low-energy subspace. 
The parameter $g_{\orb}$ controls the strength of the orbital depairing effect and is layer-number-dependent in multilayer graphene \cite{Slizovskiy2019-wo} and TMDs \cite{Badih2025}. A schematic depiction of Fermi surfaces given by the above model corresponding to one valley (e.g. $\xi=1$) is shown in Fig.\ref{fig1 singlet}a, inset.

We now introduce pairing in the normal state arising from the above Hamiltonian. We assume that attraction is developed in the Cooper channel with either a spin-singlet or a spin-triplet gap structure, and that the gap functions are momentum-independent. Such zero-momentum pairing is possible in graphene or TMD-based systems due to the valley degree of freedom and provided that no inversion symmetry breaking occurs. The linearized gap equation for our model reads
\begin{equation}
\label{eq:linearized_gap}
 \hat{\Delta} = g_{SC} T\sum_{\omega_{n}}\int d^{2}\bs{k}G^{0}\left(\bs{k},i\omega_{n}\right)\hat{\Delta}G^{0*}\left(-\bs{k},-i\omega_{n}\right),
 \end{equation}
 where $g_{SC}>0$ -- is the coupling constant, and the gap function follows the conventional notation \cite{SigristUeda}
 \begin{equation}
 \hat{\Delta}=\left(d_0s_{0}+\bs{d}\cdot\bs{s}\right)is_y,
\label{singlet and triplet component}
\end{equation}
with $d_0$ and $\bs{d}$ parameterizing magnitude of spin-singlet and spin-triplet components correspondingly.
For simplicity, in contrast to recent microscopic theories \cite{MarylandPhonons2022PRB,Jimeno-PozoPRB2023,Cea2023PRBSolo,Zhiyu2023PRB1,Zhiyu2023PRB2,Zhiyu2023PNAS,Shavit2023PRBWeizmann,Dong2025Canting,Oxford2024PRB,Raines2025superconductivitySOC,Raines2025superconductivityMagnon,Cornell2025PRB,Sedov2025Arxiv}, here we remain agnostic about the microscopic origins of the attraction in the Copper channel and treat it phenomenologically. 
For the Hamiltonian of Eq. \eqref{reduced_free_particle_h}, the free fermion Green's functions take the form
 \begin{eqnarray}
     G_{0}\left(\bs{k}+\bs{k_0},i\omega_{n}\right) &=& \frac{\left(i\omega_{n}-z_{+}\right)s_{0}+\bP_{+}\cdot\bs{s}}{\left(i\omega_{n}-z_{+}\right)^{2}-\left| \bP_{+} \right|^{2}}, \\
      G_{0}^{*}\left(-\bs{k}-\bs{k_0},i\omega_{n}\right) &=& \frac{\left(-i\omega_{n}-z_{-}\right)s_{0}+\bP_{-}\cdot\bs{s}^{*}}{\left(-i\omega_{n}-z_{-}\right)^{2}- \left| \bP_{-} \right|^{2}},
 \end{eqnarray}
where $\bP_{\pm}=\pm\left(\bs{g}_{I}+\bs{g}_{R}\right)+\frac{1}{2}g\mu_B\bs{B}$ defines the effective ``magnetic'' field experienced by the electrons in the two valleys (which form the Cooper pair with zero total momentum) and $z_\xi=\epsilon_\bs{k}-\xi z_0$. Here for clarity of notation in the Green's function we measure momentum $\bs{k}$ from the trigonal warping pockets $\bs{k}_0$ in the two valleys ($\xi$), c.f. Fig.\ref{fig1 singlet}a, inset. 

The linearized gap equation \eqref{eq:linearized_gap} allows us to readily calculate the upper critical magnetic field for a given symmetry of the superconducting gap function. In our calculations we assume that neither the density of electron states (DOS) at the Fermi level $N(E_F)$, nor the topology of the Fermi surface and the area inside it change significantly. In the calculation that follows, allowing us to simplify the gap equation, we also assume that the Fermi energy $E_F$ is the largest energy scale (compared to SOC) in the problem - an assumption we verify when applying the model to experimental data in the last section of the paper. We also note that the assumptions made in writing Eq. \eqref{eq:linearized_gap} regarding the pairing mechanism (e.g., constant attraction) and the lack of momentum dependence in the order parameter are idealizations that are likely not true in realistic systems. For the purpose of developing a unifying framework, however, they are well-motivated, as they allow us to compare multiple systems to identify emerging trends easily. Lastly, we employ a decoupling of the spin-singlet and spin-triplet gap equations, which is justified in the limit of small SOC compared to the Fermi energy \cite{AgterbergPRL2004}. 

The advantage of such simplified structure of the linearized gap equation allows us to absorb both the coupling constant $g_{SC}$ and $N(E_F)$ into $T_{c0}$. Here $T_{c0}$ is defined as the standard BCS critical temperature in the absence of SOCs and external fields. Specifically, as we show in the SI, the equations that define the critical magnetic field $H_{c_2}(T)$ at a temperature $T$ (in our calculation given by the value of $B$ which solves the linearized gap equation) for both spin-singlet and spin-triplet gap functions can be cast into the same form:
\begin{widetext}
\begin{equation}
\begin{gathered}
\ln \left(\frac{T}{T_{c1}}\right)+ \langle\Phi(\rho_{+},Z_{0})+\Phi(\rho_{-},Z_{0})-\chi[\Phi(\rho_{-},Z_{0})-\Phi(\rho_{+},Z_{0})]\rangle_{FS}=\langle(1+\chi_{B=0})\Phi(\rho_{+}|_{B=0,T=T_{c1}},0)\rangle_{FS},
\end{gathered}
\label{general_gap_eq}
\end{equation}
\end{widetext}
where
\begin{eqnarray}
    &&\Phi(\rho,Z_{0})\equiv\frac{1}{4}\text{Re} \left[\Psi \left( \frac{1+i\rho}{2}+iZ_{0} \right)  \right. \nonumber \\ 
    &&\left. +\Psi \left( \frac{1+i\rho}{2}-iZ_{0} \right) - 2\Psi \left( \frac{1}{2} \right) \right], \nonumber \\
    &&\rho_{\pm}\equiv \frac{|\bP_{+}|\pm|\bP_{-}|}{2\pi T},\quad Z_{0}\equiv \frac{z_{0} }{2\pi T}, \nonumber
\end{eqnarray}
with
\begin{equation}
    \chi\equiv 
    \begin{cases} 
        \frac{\textbf{P}_{+}\cdot\textbf{P}_{-}}{|\bP_{+}||\bP_{-}|}, & \text{for spin-singlet} \\
        -\frac{\textbf{P}_{+}\cdot\textbf{P}_{-}}{|\bP_{+}||\bP_{-}|}+\frac{2(\textbf{P}_{+}\cdot \textbf{d})(\textbf{P}_{-}\cdot \textbf{d})}{|\bP_{+}||\bP_{-}| \left| \textbf{d} \right|^2}, & \text{for spin-triplet}
\end{cases}
\nonumber
\end{equation}
and $\Psi(x)$ -- is the digamma function. Depairing in the system is controlled by the parameter $\rho$ in our definition of $\Phi (\rho, Z_0)$. For clean systems, $\rho$ enters as the imaginary part of the digamma function argument. In contrast, in the dirty limit, where depairing is mainly due to spin-orbital scattering and magnetic impurities, $\rho$ enters as a real part \cite{FischerHelvetica1973, Tinkham}. Here the notation
$\langle\cdots\rangle_{FS}$ indicates averaging over the Fermi surface, which can become non-trivial in the presence of $\textbf{k}$-dependent Rashba SOC. 

The different forms of $\chi$ for spin-singlet and spin-triplet cases are from the trace over the spin space. $\chi\in [-1,1]$, and its specific value encodes the misalignment of the spins of two electrons forming the Cooper pair: the smaller $\chi$ is, the easier it is for the two electrons to get paired. The $T_{c1}$ is defined as the critical temperature in the absence of a magnetic field ($B=0$), but in the presence of SOCs; it is, in principle, the critical temperature which can be measured in experiment - a fact to which we come back in the final part of the paper. For the singlet channel $T_{c1}=T_{c0}$ since the spin-singlet case has $\chi_{B=0}=-1$, which is consistent with the earlier work of \cite{AgterbergPRL2004}. For the spin-triplet case $\chi_{B=0}$ depends on SOC and $T_{c1}$. In the case of spin-triplet pairing channel, a nonzero right-hand side of Eq. \eqref{general_gap_eq} makes the critical temperature $T_{c1}$ lower than the one without SOC ($T_{c0}$) - again in agreement with Ref. \cite{AgterbergPRL2004}.

\section{Spin-singlet pairing}

The solution for the upper critical field at a particular critical temperature $T_{c}$ is defined by the nonlinear implicit equation, i.e., 
Eq. \eqref{general_gap_eq}. The nonlinear form of the equation, in particular the presence of digamma functions, makes it challenging to analytically study the $H_{c2}(T)$ behavior for intermediate temperatures. Most experimental measurements, however, focus on the vicinity of $T_{c1}$, the critical temperature at zero field, as sweeping the temperature continuously from 0 to $T_{c1}$ is challenging. Theoretically, it is also informative to consider the $T\to 0$ limit. Both of these limits can be carried out analytically, as we do below. However, for careful accounting of experimental data over a wide temperature range, we can evaluate the Fermi surface averages in Eq. \eqref{general_gap_eq} to obtain a nonlinear equation in $4$ parameters. The resulting fitting problem is then a nonlinear one that can be solved using standard numerical methods -- see SI for more discussion.

The overall qualitative trends are shown in Fig. \ref{fig1 singlet} for a numerical solution of the Eq. \eqref{general_gap_eq}.
We find that any amount of Ising SOC increases the Pauli limit violation ratio, $\text{PVR} = \frac{B}{B_p}$ with $B_P=1.76k_BT_{c1}/(\sqrt{2}\mu_B)$, at all temperatures. Moreover, in the strict limit of zero Rashba SOC, zero orbital coupling, and Ising SOC larger that the superconducting gap (set by the scale of $T_{c0}$ yields a diverging PVR. Specifically, close to zero temperature, the gap equation can be expanded to arrive at a scaling  $H_{c2}\rightarrow g^{-1}\left(T\ln\left(\lambda_I/T\right)\right)^{-1/2}$ (see Ref. \cite{Ilic2017} and the discussion in SI), confirming the numerical results. Upon introduction of a finite Rashba SOC that gives rise to depairing, the PVR at $T_c\rightarrow0$ becomes finite, with the specific intercept value approximated by 
\begin{equation}
\text{PVR}=\frac{\lambda_I}{gB_P}\left(\frac{g^2\lambda_R^2k_F^2}{4g_{\orb}^2\lambda_I^2}-1\right)^{\frac{1}{2}}=\frac{\sqrt{2}e^\gamma\lambda_I}{\pi gk_BT_{c0}}\left(\frac{g^2\lambda_R^2k_F^2}{4g_{\orb}^2\lambda_I^2}-1\right)^{\frac{1}{2}}
\end{equation}
for $B \ll T_{c1}$, specifically when $\lambda_R\ll\lambda_I$.

Near the critical temperature $T_{c1}$, the Eq. \eqref{general_gap_eq} can be expanded in powers of the in-plane magnetic field $B$ (technically in powers of $B$ over the gap) and deviation of critical temperature $(T_C(B)-T_{c1})/T_{c1}$ to give
\begin{equation}
T_c (B) \simeq T_{c1} - c_s B^2,
\label{Tc_expansion1}
\end{equation}
similar to the results from the Ginzburg-Landau theory \cite{Tinkham, MatsuokaPRR2020}. A slightly more convenient representation of Eq. \eqref{Tc_expansion1} involves PVR and reads  
\begin{equation}
T_c /T_{c1} \simeq 1 - \tilde{c}_s \text{PVR}^2.
\label{PVReq}
\end{equation}
 In our theory we calculate $\tilde{c}_s$ by expanding each term in Eq. \eqref{general_gap_eq} in small $B\rightarrow 0$ and $\delta\rightarrow0$, where $T_c/T_{c1} = 1 - \delta$ (see SI for details) to arrive at:
\begin{equation}
    \begin{gathered}
        \tilde{c}_s
=a_0\frac{(2\lambda^{2}_{I}+k^{2}_{F}\lambda^{2}_{R})g^2k_B^2 T^2_{c1}}{(\lambda^{2}_{I}+k^{2}_{F}\lambda^{2}_{R})^{2}}\Phi \left(\frac{\sqrt{\lambda^{2}_{I}+k^{2}_{F}\lambda^{2}_{R}}}{2\pi T_{c1}},0 \right)\\
        +a_1 \left[ \frac{g^2k^{2}_{F}\lambda^{2}_{R}}{8(\lambda^{2}_{I}+k^{2}_{F}\lambda^{2}_{R})}+g_{\orb}^2 \right].
    \end{gathered}
    \label{cs_PRV}
\end{equation}

\begin{table}[t]
    \centering
    \begin{tabular}{c|c|c|c}
         &  spin-$z$ basis&  spin-$x$ basis& spin-$y$ basis\\
         \hline
 $d_0$& $is_y$& $is_x$&$is_z$\\
         $d_x$&  $is_xs_y=-s_z$&  $is_zs_x$& $is_ys_z$\\
         $d_y$&  $is_y^2$&  $is_x^2$& $is_z^2$\\
         $d_z$&  $is_zs_y$&  $is_ys_x$& $is_xs_z$\\
          $\lambda_I$&  $\xi s_z$&  $\xi s_y$& $\xi s_x$\\
         $\lambda_R$&  $s_x,\;s_y$&  $s_z,\;s_x$& $s_y,\;s_z$\\
         $B$&  $s_x$&  $s_z$& $s_y$\\
    \end{tabular}
    \caption{Matrices of order parameters and spin-polarized terms in different basis.}
    \label{tab:spin-basis}
\end{table}

Here $a_0=1.56=\pi^2e^{-2\gamma}/2$ and $a_1=0.33=-e^{-2\gamma}\psi^{(2)}\left(\frac{1}{2}\right)/16$ is numerical constant related to the digamma function, with $\gamma$ as Euler constant and $\gamma=-\psi(\frac{1}{2})-2\ln 2$. As we see from Eq. \eqref{cs_PRV}, the presence of SOC decreases $\tilde{c}_s$ and leads to the enhancement of the upper critical field $H_{c2}$, in agreement with the trends of Fig. \ref{fig1 singlet}. In the specific limit of $\lambda_I\gg \lambda_R,g_{\orb}\mu_BB,T_{c1}$, we arrive at
\begin{equation}
\begin{gathered}
 \tilde{c}_s=\frac{2a_0g^2k_B^2 T^2_{c1}}{\lambda^{2}_{I}}\Phi \left(\frac{\lambda_{I}}{2\pi T_{c1}},0 \right)+a_1g_{\orb}^2 
 +\frac{a_1g^2}{8}\frac{k_F^2\lambda^{2}_{R}}{\lambda^{2}_{I}}\\
 +\frac{2a_0g^2k_B^2 T^2_{c1}}{\lambda^{2}_{I}}\left. \left[\frac{\rho}{2}\Phi'(\rho,0) -\Phi \left(\rho,0 \right)\right]\right|_{\rho=\frac{\lambda_I}{2\pi T_{c1}}}\frac{k^{2}_{F}\lambda^{2}_{R}}{\lambda^{2}_{I}}  
\end{gathered}
\end{equation}
which limit of $\lambda_R=g_{\orb}=0$ gives $\tilde{c}_s=2a_0g^2k_B^2 T^2_{c1}\Phi \left(\frac{\lambda_{I}}{2\pi T_{c1}},0 \right)/\lambda^{2}_{I}$ Further expanding the digamma function gives the leading term as $\ln(\frac{\lambda_{I}}{2\pi T_{c1}})/\lambda_I^2$ in agreement with the Refs. \cite{KTLawIsingScience2015,Ilic2017} and yielding a large PVR already at $T\to T_{c1}$. The above equation also defines a characteristic scale for $\lambda_R,g_{\orb}, T_{c1}$ above which depairing due to these mechanisms alters the pure Ising limit (See SI for further discussion).

\section{Spin-triplet pairing}

\begin{figure*}[t]
    \centering
    \includegraphics[width=\linewidth]{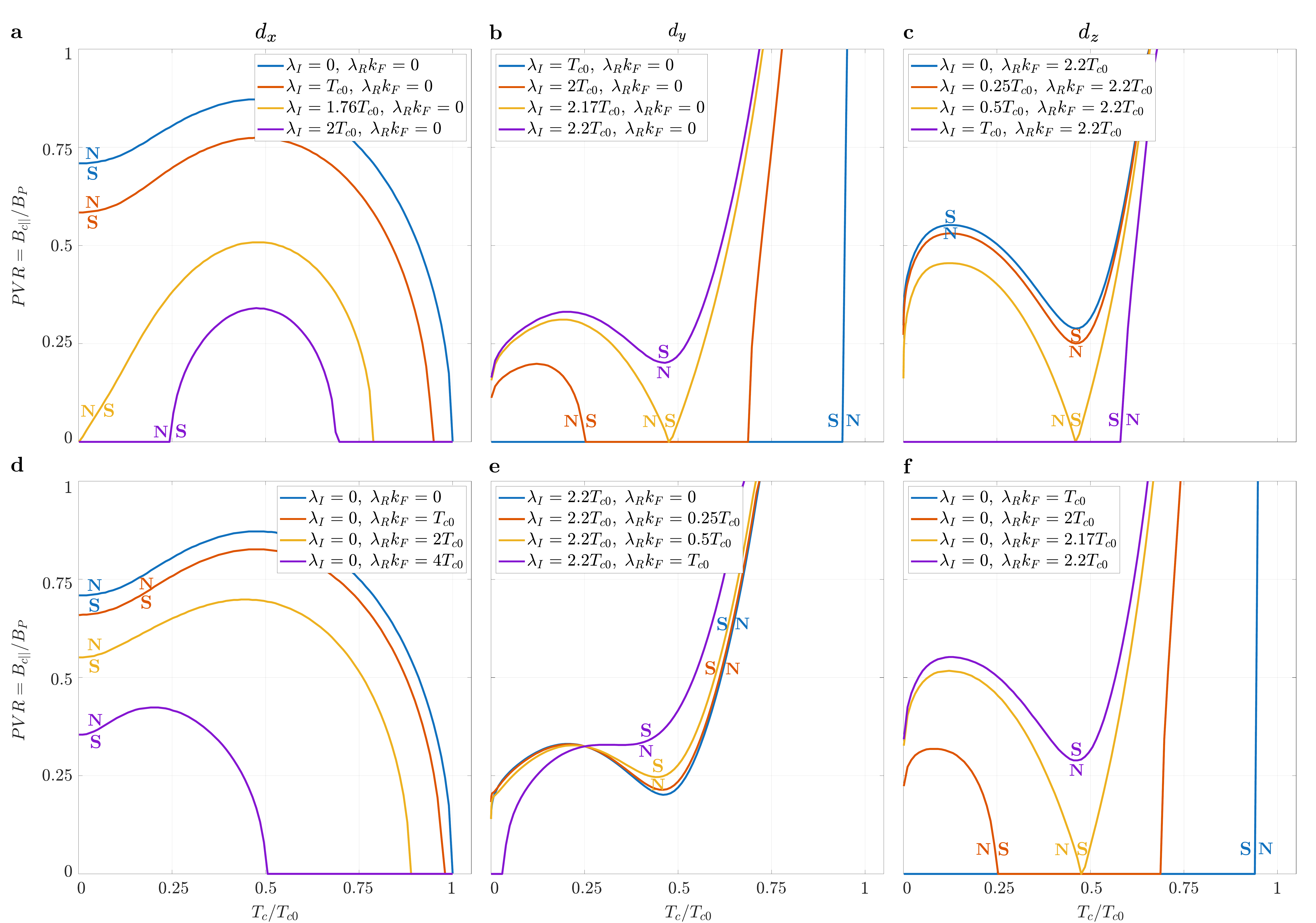}
    \caption{\textbf{a}-\textbf{c}: The upper critical field in spin-triplet superconductor with varying Ising SOC when order parameter along x, y, and z direction; \textbf{d}-\textbf{f}: The upper critical field in spin-triplet superconductor with varying Rashba SOC when order parameter along x, y, and z direction. In \textbf{a}-\textbf{f} the superconducting and normal regions as defined by the critical PVR curve are denoted with ``S'' and ``N'' letters respectively. The color of the letters corresponds to the relevant curve shown in the panel.}
    \label{fig2}
\end{figure*}

Now we consider the case of spin-triplet pairing. Fig. \ref{fig2}a-f shows the upper critical field dependence for three different orientations of the order parameter $\bs{d}$: $d_x, d_y, d_z$. Generally, the order parameter can be any linear combination of these terms, and we anticipate the curve to smoothly interpolate between the different behaviors shown in Fig. \ref{fig2} (see SI for further discussion). We find that the behavior of the critical curve $H_{c2}(T)$ qualitatively differs for the order parameter $d_x$ (Fig. \ref{fig2}a,d) compared to that of $d_y,d_z$ (Fig. \ref{fig2}b,c,e,f). This difference arises from the orientation of the in-plane magnetic field that is set along the $x$-axis in our model. For the $d_x$ parameter, the behavior qualitatively resembles that of the spin-singlet pairing in the absence of SOCs, which can be qualitatively understood from the anti-commutation of the Pauli matrices $\left\{s_x,s_{y,z}\right\}=0$. However, once Ising SOC is introduced, its role is different from spin-singlet as the corresponding anti-commutator of the order parameter and Ising SOC differ. As a result, an increase of Ising SOC induces depairing for the $d_x$, unlike in the spin-singlet case. Generally, in the absence of SOCs,the spin-triplet pairing with order parameter along the magnetic field has the same $H_{c2}$ curves with the spin-singlet pairing. This observation is nicely demonstrated by rewriting the order parameters and SOC couplings in the different spin basis as our effective model of Eq. \eqref{reduced_free_particle_h} is written in the spin oriented along the $z$-axis basis - see Table \ref{tab:spin-basis}. For example, in the absence of SOC, the $d_x$ order parameter has an $i s_y$ representation in the spin-$x$ basis mirroring the $i s_y$ representation of the spin singlet in the spin-$z$ basis. Mathematically this mapping between spin-singlet and spin-triplet behavior is also demonstrated in the gap equation of Eq. \eqref{general_gap_eq} through the parameter $\chi$. If $\chi$ is the same for two given order parameters (say spin-singlet and $d_x$) and SOC values, then the resulting $H_{c2}(T)$ curves will be identical.

The cases of $d_y$ and $d_z$ spin-triplet order parameters provide an even richer response to in-plane magnetic fields - Fig. \ref{fig2}b,c,e,f. We want to pause here to clarify how to interpret Fig. \ref{fig2}: the PVR curves mark the boundary between the normal and superconducting states. In the spin-singlet or $d_x$ triplet case, the identification of which region is superconducting is more straightforward; however, in the $d_y$ and $d_z$ cases, the curves require careful consideration (see also SI for more discussion) to indicate which region is superconducting. In general, we find that a finite magnetic field is needed to overcome the Ising or Rashba-induced spin splitting and cant the spins into a spin structure favorable to spin-triplet pairing. The curves then exhibit non-monotonic behavior, with finite Rashba SOC pushing the onset of in-plane-magnetic-field-induced superconductivity to higher fields. In our calculation, such superconductors then appear to persist even as $B \to \infty$. However, in realistic systems, there would be an upper critical field set by the breakdown of the approximations made, e.g., when Fermi energy stops being the largest energy scale or substantial mixing between spin-singlet and triplet components develops. The $d_y$ and $d_z$ components exhibit different behavior in response to finite SOC coupling. In fact, the cases of Ising and Rashba SOC qualitatively map onto each other, as shown in Fig.~\ref {fig2}b,f. This behavior stems from the interplay between the spin-triplet decomposition basis, in which the $d_z$ channel is taken out of the plane, much as in the Ising SOC coupling. In both cases the order-parameter is perpendicular to the plane spanned by the magnetic field and the relevant SOC. In turn this leads to identical anti-commutation of the Pauli matrices, resulting in the PVR curves and appropriate SOC scales in the two cases. In the SI, we substantiate the above qualitative reasoning by analytically expanding the triplet components for $T\to 0$ and $T\to T_{c1}$.

Finally, we highlight that in the figure, we normalize the temperature axis by a quantity $T_{c0}$ to facilitate easy comparison between figures. This temperature scale however is a purely theoretical quantity that corresponds to the would-be temperature scale for a singlet pairing. The actual experimental temperature, $T_{c1}$, is in fact suppressed by the SOC coupling in the spin-triplet case. This suppression can be inferred from Eq. \eqref{general_gap_eq}, which, in the presence of SOC, has a nonzero constant RHS that exponentially decreases the critical temperature (see SI for more information).

\section{Derivation of the effective low-energy model for Bernal bilayer graphene}

Here we relate the effective model of Eq. \eqref{reduced_free_particle_h} to a microscopic Hamiltonian of Bernal bilayer graphene (BBG) with substrate-induced SOC. The continuum Hamiltonian of a microscopic model of BBG with proximity-induced SOC coupling takes the form \cite{McCann_2013_bilayer_graphene,  Rachet, Slizovskiy2019-wo, Maryland2022PRB, Jimeno-PozoPRB2023, SpinH, Zhumagulov2024PRB, CaltechTheoryPRB2024}
\begin{equation}
    \begin{gathered}
        \mathcal{H}_{\xi} (\bs{k}) = 
        \begin{pmatrix}
        -\frac{U}{2} & v \pi_1^{\dagger} & -v_4 \pi^{\dagger} & -v_3 \pi \\
        v \pi_1 & -\frac{U}{2}  & \gamma_1 & -v_4 \pi^{\dagger} \\
        -v_4 \pi & \gamma_1 & \frac{U}{2} & v \pi_2^{\dagger} \\
        -v_3 \pi^{\dagger} & -v_4 \pi & v \pi_2 & \frac{U}{2} 
        \end{pmatrix}  s_0\\
       +\frac{1}{2}g_0 \mu Bl_0\sigma_{0}s_{x} +\frac{\xi}{2}\begin{pmatrix}
            \lambda_I^{1} & 0 \\
            0 & \lambda_I^{2}
        \end{pmatrix}\s_0s_z\\ 
        + \frac{1}{2}\begin{pmatrix}
            \lambda_R^{1} & 0 \\
            0 & \lambda_R^{2}
        \end{pmatrix}\left( \xi \sigma_x s_y - \sigma_y s_x \right)\,,
    \end{gathered}
\end{equation}
 where $\lambda_I^{(l)},\lambda_R^{(l)}$ are the magnitudes of substrate-induced Ising and Rashba SOCs in the corresponding layers $l=1,2$. Here we take the g-factor $g_0=2$ in the Zeeman coupling. Here $\pi = \xi p_x + i p_y$, $\pi_1 = \xi p_x + i\left(p_y+edB/2\right)$, $\pi_2 = \xi p_x + i\left(p_y-edB/2\right)$ represent the kinetic momentum in the presence of the magnetic field ($d$ is the BBG thickness), which induce the inter-layer orbital effect, and $\sigma_i,s_i$ -- are Pauli matrices acting in the sublattice and spin space respectively. The Ising and Rashba SOCs in BBG are induced by proximity to the WSe$_2$ substrate \cite{PhysRevB.92.155403, PhysRevLett.119.146401, Wang2015, Yang2016, PhysRevB.96.041409, PhysRevB.96.125405, PhysRevLett.120.106802, PhysRevB.97.075434, Avsar2014, Dankert2017, Ghiasi2017, PhysRevB.97.045414, Bentez2017,PhysRevX.6.041020,Island2019_BBG_WSe2_proximity}. Without loss of generality, we thus assume that the SOCs on the top layer $\lambda^{l=2}_{I}=\lambda^{l=2}_{R}=0$, while the bottom layer SOC parameters are nonzero. The choice of the SOC layer structure is reflected in the sign of the external layer potential $U$, which then polarizes charges away from and towards the interfaces appropriately, e.g. for $U<0$ conduction band is polarized towards the non-zero SOC layer $l=1$. We note that we do not account here for a self-consistent screening of the external displacement field $U$ by the electronic charges \cite{PhysRevB.81.125304,PhysRevB.75.235433,Kolar2023, Kolar2025} - an effect that we anticipate to not be crucial in the bilayer system. 
 
 To obtain a low-energy theory that maps onto the $2\times 2$ spin basis of our effective Hamiltonian of Eq. \eqref{reduced_free_particle_h}, we carry out a systematic expansion treating $v$ ($\gamma_0$) and $\gamma_1$ as large parameters, followed by an expansion in powers of the SOC coupling as small parameters (e.g. see \cite{McCann_2013_bilayer_graphene, 2019arXiv190101294Z}). The difference in our analysis compared to that of Ref. \citenum{McCann_2013_bilayer_graphene, 2019arXiv190101294Z} is that we keep both the trigonal warping term, $v_3$, which is necessary to account for the three pockets as well as the displacement field $U$ necessary to polarize charges to one side. Moreover, we carry out the momentum expansion around the trigonal warping pockets centers. 
 
 Results of this mapping to the effective SOC parameters of the Eq. \eqref{reduced_free_particle_h} are shown in the equations below and the details of the derivation are provided in the SI. The behavior of the Eq. \eqref{reduced_free_particle_h} effective parameters is, in general, a non-trivial function of all the model parameters, as expected from perturbation theory analysis. The effective form of Ising and Rashba SOC, orbital effect, and Zeeman terms in the $2 \times2$ spin-basis projected to  conduction/valence band (in the spin up and down basis) are:
\begin{align}
 h_{\xi} \left(\bs{k}\right) &= \left( \epsilon_{\xi}\left(\bs{k}\right)-z_{0} \right)s_{0} + \left( \bs{g}_{I}+\bs{g}_{R}+\bs{b} \right) \cdot \bs{s}\\
z_0 &\approx \frac{Uedv^2 k_y}{2\gamma^2_1} B\label{eq:eff_orbital}\\
\bs{g}_{I}\cdot \bs{s} &\approx \frac{1}{2} \xi \lambda_I^{l=1}  s_z\\
\bs{g}_{R}\cdot \bs{s} &\approx \frac{1}{2} \left(g_{R,y}(\bs{k}) s_y-g_{R,x}(\bs{k}) s_x\right)\label{eq:effective_rashba}\\
\bs{b}\cdot \bs{s} &\approx \frac{1}{2}g \mu_B B s_x\,,
\end{align}
where we define the functions $g_{R,x}(\bs{k}), ~g_{R,y}(\bs{k})$ and $g$ below. Here we expanded around a trigonal warping center $\bs{k}_0= k_0\hat{x}$ located along the $x$-axis, c.f. Fig. \ref{fig1 singlet}a. The dispersion of other pockets can be analogously derived by rotating $\bs{k}_0$. The scale $k_0$ is given by:
\begin{equation}
    k_0 = \frac{\sqrt{\gamma_1^2 v_3^2+4 U^2 v^2}+3 \gamma_1 v_3}{4 v^2}
\end{equation}
We highlight here that as the displacement field grows, the trigonal warping centers move away from the $K, K'$ points with $U v$ actually becoming comparable to $\gamma_1 v_3$ for the relevant $U$ in the experiments (see also discussion in Ref. \cite{Zhang2025-ty}). The effective Rashba couplings $g_{R,x}(\bs{k})$ and $g_{R,y}(\bs{k})$ are 
given by:
\begin{align}
\label{effective_rashba}
g_{R,x}(\bs{k}) &\approx \lambda_R^{l=1} \frac{k_y v \left(14 \gamma_1 k_0 v_3+U^2\right)}{2 \gamma_1^2 \left|U\right|}\\
g_{R,y}(\bs{k}) &\approx \lambda_R^{l=1} \left[ \frac{\xi k_0 \left|U\right| v}{2 \gamma_1^2} +\frac{k_x  v \left(10 \gamma_1 k_0 v_3+3 U^2\right)}{2 \gamma_1^2 \left|U\right|}\right]\,
\end{align}
We find that there are two 
contributions to the effective Rashba coupling: ($i$) one set by the location of the 
trigonal warping pocket $\bs{k}_0$ and ($ii$) one controlled by the doping of each pocket 
and hence momentum $k_x,k_y$. Physically, the two origins of the terms make 
sense: ($i$) controls the winding of the Rashba spin texture around the $K, K'$ points, 
and ($ii$) controls the winding of the Rashba spin texture around the trigonal warping 
center. Finally the $g_{\eff}$ takes the form ($g_0 = 2$)
\begin{equation}
\label{eq:effective_g}
    g \approx g_0 ~(1+\frac{v^2 k_0^2}{\gamma^2_1})
\end{equation} 
In all the above expression, we focused on the leading order terms for typical 
numerical values of the microscopic model parameters.

Crucially, from the above result, we find that the effective Ising in each trigonal 
pocket is, to leading order, unmodified from the microscopic Ising value by the 
projection to each pocket. However, the effective orbital, Rashba and Zeeman contributions 
are, as a result of the projection, modified from their bare values by the applied gate voltage (external layer potential difference $U$). The effect of $U$-driven renormalization acts to increase the effective and Zeeman  couplings, but decreases the effective Rashba SOC. The inversion symmetry breaking from the displacement field and one-side substrate leads to an asymmetry between the effective SOC parameters in the conduction bands, as well as possible mixing between the two sets of bands. We provide the full effective low-energy Hamiltonian in the conduction-valence spin up-down basis in the SI.

\section{Comparison with experimental data}

We apply our developed methodology to gain insights into the properties superconductivity in BBG. In Fig. \ref{fig1 singlet}c, we show the upper critical field data extracted from Refs. \cite{Zhang2023_Enhanced_SC_proximitized_BBG, Zhang2025-ty, Holleis2025_Nemacity_BBG, Tingxin_Li2024_BBG_WSe2}. These data sets correspond to the so-called SC2 phase, which is believed to be an SC formed from a subset of the trigonal warping pockets. Within our framework, however, the precise number of occupied pockets is irrelevant; it is absorbed into the definition of $T_{c0}$ and the precise parent-state Fermi surface would manifest if the direction of the in-plane field were to be varied (see Ref. \cite{Zhang2025-ty} for more discussion).  

The summary of fitted parameters is shown in Table \ref{tab:fitting result}. From our fitting, we find that the effective Rashba SOC is negligible, unlike the anticipated bare-layer-induced Rashba on the scale of several meV  (e.g. Ref. \cite{Island2019_BBG_WSe2_proximity}). Such a fitting result is consistent with the result of the effective model, which reduces the effective Rashba SOC, see Eq. \eqref{effective_rashba}. The fitted Ising SOC values are slightly smaller than, but of comparable magnitude to the values reported from the Landau level crossing. This observation is in line with the theoretically anticipated lack of renormalization of the Ising SOC. The fitted value of the orbital coupling is close to the theoretically expected values of $g_{\orb}$ ranging from 0.18 to 0.25 (recall that $g_{\orb}$ is also $U$-dependent, c.f. Eq. \eqref{eq:eff_orbital}). 

Interestingly, however, we find that the resulting $g$-factor values are greater than 2. We conjecture that this renormalization for the $g$ factor may be due to a possible interaction enhancement of the Zeeman energy scale \cite{LarentisMoSe2} as the ``non-interacting'' enhancement of the effective $g$ factor due to the projection process, Eq. \eqref{eq:effective_g}, gives an enhancement by at most $\sim 8\%$. We do, however, note that it is also feasible that our theory overestimates the $g$-factor correction, as the critical temperature $T_{c1}$ in our model is the BCS critical temperature (i.e., the pair-formation temperature scale). In contrast, experimentally, it is likely to be set by the BKT transition temperature, which is lower than the BCS value that could yield a smaller, in line with the $g=2$ expectation, fitting result. In the SI, we provide other fitting models that allow for different behavior of the SOC constants across the samples. We confirmed that the fitted values of all the parameters validated our model's assumptions, in particular the Fermi energy is the largest energy scale in the problem around $E_F \sim 30$ meV.

\begin{table}[htbp]
    \centering
    \begin{tabular}{c|cccccccc}
         &  n &  U  &$T_{c0}$ &  reported &$\lambda_I$ &  $\lambda_R$ &  $g_{\orb}$& $g/g_0$\\
         & $(10^{11} cm^{-2})$ & (meV) & (K) &$\lambda_I$ (meV)&(meV) & (meV) & & \\
        \hline
         \textbf{a}&  -7.0&  110& 0.28& 0.7&0.53&  0.09&  0.13& 3.46\\
         \textbf{b}&  -7.3& 120& 0.39& 1.5&1.38&  0.15& 0.09& 2.52\\
         \textbf{c}&  -7.3&  115 &0.28&  1.6&1.05&  0.03&  0.15& 2.83\\
         \textbf{d}&  -5.9&  96 &0.19&  1.7&0.38&  0.01&  0.16& 2.56\\
    \end{tabular}
    \caption{Fitting results of the upper critical field in SC2 phase in BBG/WSe$_2$. The fitting is generated by varying $\lambda_I$, $\lambda_R$, $g_{\orb}$, and Land\'e $g$. The experimental data were extracted from \textbf{a}\cite{Zhang2023_Enhanced_SC_proximitized_BBG}, \textbf{b}\cite{Zhang2025-ty}, \textbf{c}\cite{Holleis2025_Nemacity_BBG}, and {\textbf{d}}\cite{Tingxin_Li2024_BBG_WSe2}. }
    \label{tab:fitting result}
\end{table}

\section{Discussion and conclusions}

In this work, we developed a general procedure to calculate the upper-critical field of a clean 2D electron system in the presence of Ising and Rashba SOC, the in-plane orbital effect, and the Zeeman effect for either spin-singlet or spin-triplet superconducting pairing state. While we specifically focused on Bernal bilayer graphene as a motivating experimental platform, our results apply to other experimental systems as well, provided that a mapping to the effective Hamiltonian of Eq. \eqref{reduced_free_particle_h} can be established. Our model is a generalization of the existing works on the SOC-driven depairing of superconductivity \cite{EdelsteinPRL1995, GorkovRashbaPRL2001, GorkovBarzykin2002, AgterbergPRL2004, SergienkoPRB2004, AgterbergVorticesPRL2005, DimitrovaPRB2007, KTLawIsingScience2015, Smidman_2017, FuldeJPSJ2017, Ilic2017, ShafferPRB2020, JinPRB2025, WangPRL2019-TypeIIIsing}. Its distinguishing feature is the separation of the different SOC parameters in the analysis and discussion of how they affect the singlet and triplet components, both numerically over the full temperature range and analytically in the high- and low-temperature limits.

We used our framework to extract the values of Rashba and Ising SOC, the orbital and Zeeman couplings from experimental upper critical in-plane magnetic field measurements in Bernal bilayer graphene proximitized with WSe$_2$ \cite{Zhang2023_Enhanced_SC_proximitized_BBG,Zhang2025-ty, Holleis2025_Nemacity_BBG,Tingxin_Li2024_BBG_WSe2}. We found that while the values of Rashba, Ising SOC, and orbital coupling are in line with previously reported and theoretically expected values, the $g$-factor consistently appears to be enhanced compared to the standard value of 2 - a value that cannot be accounted for by a ``non-interacting''  renormalization of the effective $g$ factor due to finite displacement fields present in experiments.

Due to the simplicity of the formalism, our theory is based on multiple assumptions. These are: (i) superconducting pairing being a single, momentum-independent coupling constant, (ii) Fermi energy being the largest energy scale in the problem, (iii) the transition out of the superconducting state being a second-order one, and (iv) interpretation of the experimental $T_c$ as the critical temperature in our theory. Each of these approximations can be relaxed, and then it is interesting to analyze how the expected behavior of the critical in-plane magnetic fields would be modified. 

\section{Acknowledgments}

H.M. and C.L. are supported by start-up funds from Florida State University and the National High Magnetic Field Laboratory. D.V.C. acknowledges financial support from the National High
Magnetic Field Laboratory through a Dirac Fellowship and from Washington University in St. Louis through the Edwin Thompson
Jaynes Postdoctoral Fellowship. The National High Magnetic Field Laboratory is supported by the National Science Foundation through NSF/DMR-2128556 and the State of Florida. 

\bibliography{biblio}
\clearpage
\onecolumngrid 

\begin{center}
    \textbf{\large Supplementary information for "Upper critical in-plane magnetic field in quasi-2D layered superconductors"}\\[.2cm]
    Huiyang Ma$^{1,2}$, Dmitry V. Chichinadze$^{1,3}$, and Cyprian Lewandowski$^{1,2}$\\[.1cm]
    {\itshape \small 
    $^1$National High Magnetic Field Laboratory, Tallahassee, Florida 32310, USA\\
    $^2$Department of Physics, Florida State University, Tallahassee, Florida 32306, USA\\
    $^3$Department of Physics, Washington University in St. Louis, St. Louis, Missouri 63160, USA\\
    }
\end{center}
\vspace{8mm}

\setcounter{equation}{0}
\setcounter{figure}{0}
\setcounter{table}{0}
\setcounter{section}{0}
\renewcommand{\theequation}{S\arabic{equation}}
\renewcommand{\thefigure}{S\arabic{figure}}
\renewcommand{\thetable}{S\arabic{table}}
\renewcommand{\thesection}{S\arabic{section}}

\section{Derivation of the upper critical field $B_{c2}$}

\subsection{General formalism}

To study the formation of superconductivity and its upper-critical in-plane magnetic field temperature dependence, we assume that the effective potential is attractive and constant to momentum $\bs{p}$ near the K (K') points. We exclude the FFLO state from consideration and ignore the reshaping of the Fermi pocket by the external magnetic field and SOC. We also assume no inter-band pairing between valence and conduction bands. The mixing of spin-singlet and spin-triplet pairing is in the order of $O\left(\lambda_{I,R}/E_F\right)$ \cite{Fischer2023_Local_Inversion-Symmetry_Breaking}, so we discuss singlet and triplet pairing separately. In our calculations for the spin-triplet pairing, we only consider the phase diagram for low magnetic field ($E_Z\ll\lambda_{I,R}$), such that the orientation of the order parameter remains unchanged. Lastly, the spin-triplet order parameter is constrained to be real, so we will not discuss chiral superconducting states.

 We consider the normal state described by a single-particle Hamiltonian $H = \sum_{\bs{k},\xi} \Psi^{\dagger}_{\bs{k},s} h_{\xi} \left(\bs{k}\right) \Psi_{\bs{k},s'}$ where
\begin{equation}
    h_{\xi} \left(\bs{k}\right) = \left( \epsilon\left(\bs{k}\right)-z_{0} \right)s_{0} + \left( \bs{g}_I+\bs{g}_{R}+\bs{b} \right) \cdot \bs{s}
    \label{reduced_free_particle_h}
\end{equation}
where $\bs{s}$ is vector of Pauli matrices acting in spin space, $g$ is the Land\'e $g$-factor to that of a free electron. Furthermore, $\bs{b} = \left(\frac{1}{2}g\mu_B B,0,0\right)$ with $B$ being the external magnetic field magnitude, $z_{0} =  g_{\text{orb}}\mu_BB\sin\theta$ 
, $\bs{g}_I = \left(0, 0, \frac{1}{2}\xi\lambda_I\right)$, $\bs{g}_R = \frac{1}{2}\lambda_R \left(k_y, -k_x, 0\right)$,  and $\xi = \pm 1$ stands for the valley index \cite{SpinH,SpinH_2,Rachet,Zhang2023_Enhanced_SC_proximitized_BBG}. Here, $\lambda_I$
and $\lambda_R$ are the Ising and the Rashba-type spin-orbit coupling, which are generically non-zero in systems with broken inversion symmetry (in graphene hetero-structures, inversion symmetry can be broken by applied external displacement field). Finally, $g_\orb$ controls the strength of the orbital depairing effect. Its magnitude depends on the number of layers in a multi-layer graphene hetero-structure \cite{Slizovskiy2019-wo}.

To calculate the upper critical field, we consider the linearized gap equation
\begin{equation}
 \hat{\Delta}= g_{SC}T\sum_{\omega_{n}}\int d^{2}\bs{k}\hat{G}^{*}_{0}\left(-\bs{k},i\omega_{n}\right)\hat{\Delta}\hat{G}_{0}\left(\bs{k},i\omega_{n}\right),
 \label{Gap_eq_general_expr}
 \end{equation}
 where
 \begin{equation}
 \hat{\Delta}=\left(d_0s_{0}+\bs{d}\cdot\bs{s}\right)is_y,
\label{singlet and triplet component}
\end{equation}
 and the free fermion Green's function takes the form
 \begin{eqnarray}
     \hat{G}_{0}\left(\bs{k}+\bs{K},i\omega_{n}\right) = \frac{\left(i\omega_{n}-z_{+}\right)s_{0}+\bP_{+}\cdot\bs{s}}{\left(i\omega_{n}-z_{+}\right)^{2}-\left|\bP_+\right|^2},\\
     \hat{G}_{0}^{*}\left(-\bs{k}-\bs{K},i\omega_{n}\right) = \frac{\left(-i\omega_{n}-z_{-}\right)s_{0}+\bP_{-}\cdot\bs{s}^{*}}{\left(-i\omega_{n}-z_{-}\right)^{2}-\left|\bP_-\right|^2}.
 \end{eqnarray}
 
Here $z_\tau=\epsilon_\bs{k}-\tau z_0$, $\bP_\tau=\tau\left(\bs{g}_{I}+\bs{g}_{R}\right)+\frac{1}{2}g\mu_B\bs{B}$, representing the effective kinetic energy and effective magnetic field in for the particle and its hole conjugate, distinguished by $\tau$. This procedure allows us to calculate the upper critical field for $\hat{\Delta}$. 
To simplify the equation, we took the density of state (DOS) $N\left(E\right)$ as a slowly varying function in the range $E-E_F\in \left[-\Lambda,\Lambda\right]$,  and ignore the change by spin splitting, $N_{\uparrow}\left(E_F\right)\approx N_{\downarrow}\left(E_F\right)=N\left(E_F\right)$.

\subsection{Spin-singlet pairing}
We start by considering a spin-singlet superconducting gap.
 For the spin-singlet pairing, the gap matrix reads  $\hat{\Delta}=\Delta is_y$. By multiplying $is_{y}$ from the left, the Eq. \eqref{Gap_eq_general_expr} becomes
\begin{equation}
 -\Delta I_{2\times 2}=g_{SC}T\sum_{\omega}\int d^2\bs{k} \;is_{y}\hat{G}^{0*}_{\omega}\left(-\bs{k}\right)is_{y}\hat{G}^{0}_{\omega}\left(\bs{k}\right)\Delta.
 \end{equation}
Taking the trace on both sides and introducing averaging over the Fermi surface, we obtain
\begin{equation}
    1 =\frac{1}{2} g_{SC}T\sum_{\omega}\int \text{tr}\left\{s_{y}\hat{G}^{0*}_{\omega}\left(-\bs{k}\right)s_{y}\hat{G}^{0}_{\omega}\left(\bs{k}\right)\right\}d^{2}\bs{k} 
    \approx\frac{1}{2}N\left(E_F\right) g_{SC}T\sum_{\omega}\int\left\langle \text{tr} \left\{s_{y}\hat{G}^{0*}_{\omega}\left(-\bs{k}\right)s_{y}\hat{G}^{0}_{\omega}\left(\bs{k}\right)\right\}\right\rangle_{FS}d\epsilon_\bs{k}.
    \label{Eq_singlet_init}
\end{equation}
Here $\left\langle\cdots\right\rangle_{FS}$ does the average over the Fermi surface, with the attendance of $k$-dependent terms. 
In this model, they are the Rashba SOC and orbital effect.  
Taking the trace yields
\begin{equation}
    \text{tr} \left\{\cdots\right\} = \frac{\text{tr}\left\{s_{y}\left[\left(-i\omega-z_{-}\right)s_{0}+\bP_{-}\cdot\bs{s}^{*}\right]s_{y}\left[\left(i\omega-z_{+}\right)s_{0}+\bP_{+}\cdot\bs{s}\right]\right\}}{\left[\left(i\omega-z_{+}\right)^{2}-\left|\bP_+\right|^2\right]\left[\left(-i\omega-z_{-}\right)^{2}-\left|\bP_-\right|^2\right]} 
    = \frac{2\left[\left(i\omega-z_{+}\right)\left(-i\omega-z_{-}\right)-\bP_{+}\cdot\bP_{-}\right]}{\left[\left(i\omega-z_{+}\right)^{2}-\left|\bP_+\right|^2\right]\left[\left(-i\omega-z_{-}\right)^{2}-\left|\bP_-\right|^2\right]}.
\end{equation}

The momentum integral of the trace now looks like 
\begin{align}
   \frac{1}{2} \int d^2\bs{k}\text{tr} \left[s_{y}\hat{G}^{0*}_{\omega}\left(-\bs{k}\right)s_{y}\hat{G}^{0}_{\omega}\left(\bs{k}\right)\right]
   &=\frac{1}{2}N\left( E_F\right) \int d\epsilon_\bs{k}\left\langle \text{tr} \left[s_{y}\hat{G}^{0*}_{\omega}\left(-\bs{k}\right)s_{y}\hat{G}^{0}_{\omega}\left(\bs{k}\right)\right] \right\rangle_{FS}\\
   &=\int d\epsilon_{\bs{k}} \left\langle\frac{\left(i\omega-z_{+}\right)\left(-i\omega-z_{-}\right)-\bP_{+}\cdot\bP_{-}}{\left[\left(i\omega-z_{+}\right)^{2}-\left|\bP_+\right|^2\right]\left[\left(-i\omega-z_{-}\right)^{2}-\left|\bP_-\right|^2\right]} \right\rangle_{FS}
\end{align}

Using the residue theorem, we evaluate the integral 
\begin{align}
    I &= \int^{+\infty}_{-\infty}d\epsilon \frac{\left(i\omega_{n}-z_{+}\right)\left(-i\omega_{n}-z_{-}\right)+a}{\left[\left(i\omega_{n}-z_{+}\right)^{2}-b^{2}\right]\left[\left(-i\omega_{n}-z_{-}\right)^{2}-c^{2}\right]}\\
    &= \int^{+\infty}_{-\infty}d\epsilon \frac{\epsilon^{2}+\left(\omega_{n}-iz_{0}\right)^{2}+a}{\left[\left(\epsilon-i\left(\omega_{n}-iz_{0}\right)\right)^{2}-b^{2}\right]\left[\left(\epsilon+i\left(\omega_{n}-iz_{0}\right)\right)^{2}-c^{2}\right]}\\
    &= \sgn\left(\omega_{n}\right)\frac{4\pi\left(\omega_{n}-iz_{0}\right)\left[b^{2}+c^{2}+2a+4\left(\omega_{n}-iz_{0}\right)^{2}\right]}{\left[\left(b+c\right)^{2}+4\left(\omega_{n}-iz_{0}\right)^{2}\right]\left[\left(b-c\right)^{2}+4\left(\omega_{n}-iz_{0}\right)^{2}\right]}.
\end{align}
Performing the Matsubara summation we obtain
\begin{equation}
    T\sum_{\omega_{n}} \left(\frac{\pi}{\left|\omega_{n}\right|}-I \right) =\Phi\left(\rho_{+},Z_{0}\right)+\Phi\left(\rho_{-},Z_{0}\right)+\frac{a}{bc}\left[\Phi\left(\rho_{-},Z_{0}\right)-\Phi\left(\rho_{+},Z_{0}\right)\right],
\end{equation}
where
\begin{equation}
    \Phi\left(\rho,Z_{0}\right)\equiv\frac{1}{4}\left \{Re \left[\Psi \left(\frac{1+i\rho}{2}+iZ_{0} \right)+\Psi \left( \frac{1+i\rho}{2}-iZ_{0} \right)-2\Psi \left( \frac{1}{2} \right) \right ] \right\},
\end{equation}
\begin{equation}
    \rho_{\pm}\equiv \frac{\left|\bP_{+}\right|\pm\left|\bP_{-}\right|}{2\pi T},\quad Z_{0}\equiv \frac{z_{0}}{2\pi T}.
\end{equation}
Here $\Psi\left(x\right)$ is digamma function defined as $\Psi\left(x\right)\equiv\frac{d}{dx} \ln\left(\Gamma\left(x\right)\right)=\Gamma'\left(x\right)/\Gamma\left(x\right).$ $\rho$ appears at the imaginary part in the clean system.
The standard BCS-like summation over Matsubara frequencies reads
\begin{equation}
   T\sum_{\omega_{n}}\frac{\pi}{\left|\omega_{n}\right|}=\sum^{\frac{\Lambda}{2\pi T}}_{n=-\frac{\Lambda}{2\pi T}}\frac{1}{\left|2n+1\right|}=\ln\left(1.13\Lambda/T\right).
\end{equation}
We now define
\begin{equation}
  \chi= -\frac{a}{bc}=\frac{\bP_{+}\cdot\bP_{-}}{\left|\bP_+\right|\left|\bP_-\right|},  
\end{equation}
where  $a=-\bP_{+}\cdot\bP_{-}$, $b=\left|\bP_+\right|$, $c=\left|\bP_-\right|$. Introduce
\begin{equation}
     F \equiv \frac{1}{ g_{SC}N\left(E_F\right)}= T\sum_{\omega_{n}} \left\langle I\right\rangle_{FS} 
      =\ln\left(1.13\Lambda/T\right)-\left\langle \Phi\left(\rho_{+},Z_{0}\right)+\Phi\left(\rho_{-},Z_{0}\right)-\chi\left[\Phi\left(\rho_{-},Z_{0}\right)-\Phi\left(\rho_{+},Z_{0}\right)\right]\right\rangle_{FS}.
     \label{Identity_in_Bc2_equation}
\end{equation}
To cancel the cutoff scale $\Lambda$, we use the identity $F\left(B\right)=F\left(B=0\right)$ and get
\begin{equation}
    \ln\left(\frac{T_{c0}}{T}\right)+\left\langle\left(1+\chi_{B=0}\right)\Phi\left(\rho_{+0},0\right)-\left\{ \Phi\left(\rho_{+},Z_{0}\right)+\Phi\left(\rho_{-},Z_{0}\right)-\chi\left[\Phi\left(\rho_{-},Z_{0}\right)-\Phi\left(\rho_{+},Z_{0}\right)\right]\right\}\right\rangle_{FS}=0
    \label{original_Bc2_equation}
\end{equation}
For $B=0$, $\chi=-1$. The $B_{c2}$ equation for spin-singlet then reads
\begin{equation}
    \ln\left(\frac{T}{T_{c0}}\right)+ \left\langle\Phi\left(\rho_{+},Z_{0}\right)+\Phi\left(\rho_{-},Z_{0}\right)-\chi\left[\Phi\left(\rho_{-},Z_{0}\right)-\Phi\left(\rho_{+},Z_{0}\right)\right]\right\rangle_{FS}=0
    \label{singlet_gap_eq}
\end{equation}
This result was obtained under the assumption of anti-symmetric spin-orbit coupling (ASOC) that $\bs{g}\left(\bs{-k}\right)=-\bs{g}\left(\bs{k}\right)$. However, the expression is similar when we replace it by the type II Ising SOC appearing in the $\Gamma$ point, which can be described by the BHZ model, and redefine $\xi$ as band index instead of valley index (see also discussion in Ref. \cite{Badih2025}).

\subsection{Spin-triplet pairing}

For the spin-triplet pairing, the gap matrix is given by  $\hat{\Delta}=\Delta \bs{d}\cdot\bs{s} is_y$, with the normalized vector $\bs{d}$.
Similarly to the spin-singlet pairing case, we multiply Eq. \eqref{Gap_eq_general_expr} by $i\bs{d}\cdot\bs{s}s_{y}$ from the left and take trace on both sides of the equation to obtain
\begin{equation}
    1 \approx\frac{1}{2}N\left(E_F\right) g_{SC}T\sum_{\omega}\int \left\langle \text{tr} \left[\bs{d}\cdot\bs{s}s_{y}\hat{G}^{0*}_{\omega}\left(-\bs{k}\right)\bs{d}\cdot\bs{s}s_{y}\hat{G}^0_{\omega}\left(\bs{k}\right) \right] \right\rangle_{FS}d\epsilon_\bs{k}.
\end{equation}
The result of the trace operation is 
\begin{align*}
    \text{tr}\left\{\cdots\right\} & = \frac{\text{tr}\left\{\bs{d}\cdot\bs{s}s_{y}\left[\left(-i\omega-z_{-}\right)s_{0}+\bP_{-}\cdot\bs{s}^{*}\right]\bs{d}\cdot\bs{s}s_{y}\left[\left(i\omega-z_{+}\right)s_{0}+\bP_{+}\cdot\bs{s}\right]\right\}}{\left[\left(i\omega-z_{+}\right)^{2}-\left|\bP_+\right|^2\right]\left[\left(-i\omega-z_{-}\right)^{2}-\left|\bP_-\right|^2\right]}\\
    & = \frac{2\left[\left(i\omega-z_{+}\right)\left(-i\omega-z_{-}\right)+\bP_{+}\cdot\bP_{-}-2\left(\bP_+\cdot \bs{d}\right)\left(\bP_-\cdot \bs{d}\right)\right]}{\left[\left(i\omega-z_{+}\right)^{2}-\left|\bP_+\right|^2\right]\left[\left(-i\omega-z_{-}\right)^{2}-\left|\bP_-\right|^2\right]}.
\end{align*}
Now the integral still has the form 
\begin{align}
    I &= \int^{+\infty}_{-\infty}d\epsilon \frac{\left(i\omega_{n}-z_{+}\right)\left(-i\omega_{n}-z_{-}\right)+a}{\left[\left(i\omega_{n}-z_{+}\right)^{2}-b^{2}\right]\left[\left(-i\omega_{n}-z_{-}\right)^{2}-c^{2}\right]}\\
    &= \sgn\left(\omega_{n}\right)\frac{4\pi\left(\omega_{n}-iz_{0}\right)\left[b^{2}+c^{2}+2a+4\left(\omega_{n}-iz_{0}\right)^{2}\right]}{\left[\left(b+c\right)^{2}+4\left(\omega_{n}-iz_{0}\right)^{2}\right]\left[\left(b-c\right)^{2}+4\left(\omega_{n}-iz_{0}\right)^{2}\right]}.
\end{align}
but now we have $a=\bP_{+}\cdot\bP_{-}-2\left(\bP_+\cdot \bs{d}\right)\left(\bP_-\cdot \bs{d}\right)/\left|\bs{d}\right|^2$. Therefore, it implies the gap equation turns out to be the same if we define 
\begin{equation}
    \chi=-\frac{\bP_{+}\cdot\bP_{-}}{\left|\bP_+\right|\left|\bP_-\right|}+\frac{2\left(\bP_{+}\cdot\bs{d}\right)\left(\bP_{-}\cdot\bs{d}\right)}{\left|\bP_+\right|\left|\bP_-\right|\left|\bs{d}\right|^2}.
    \label{triplet_chi}
\end{equation}
for spin-triplet pairing. Different form of $\chi$ came from the trace. Performing the same set of steps as in the spin-singlet case, we obtain the equation for the upper-critical field $B_{c2}$ for the spin-triplet order parameter:
\begin{equation}
\begin{gathered}
\ln \left(\frac{T}{T_{c1}}\right)+ \left\langle\Phi\left(\rho_{+},Z_{0}\right)+\Phi\left(\rho_{-},Z_{0}\right)-\chi\left[\Phi\left(\rho_{-},Z_{0}\right)-\Phi\left(\rho_{+},Z_{0}\right)\right]\right\rangle_{FS}=\left\langle\left(1+\chi_{B=0}\right)\Phi\left(\rho_{+}|_{B=0,T=T_{c1}},0\right)\right\rangle_{FS}.
\end{gathered}
\label{general_Bc2_eq}
\end{equation}
This equation reveals that the zero-field critical temperature is now SOC-dependent; we name it $T_{c1}$ and define $T_{c0}$ as the zero-field zero-SOC critical temperature. Since we assume the coupling constant $ g_{SC}$ is independent of the spin-pairing, we can get the same $T_{c0}$ for any spin-pairing form under the same DOS and coupling constant.

$T_{c1}$ and $T_{c0}$ are related in the way
\begin{align}
    0 &= F\left(B=0\right)-F\left(B=\lambda_I=\lambda_R=0\right)\\
    &= \ln\left(\frac{T_{c0}}{T_{c1}}\right)+\left\langle\left(1+\chi_{B=\lambda_I=\lambda_R=0}\right)\Phi\left(0,0\right)-\chi_{B=0}\Phi\left(\rho_{+0},0\right)\right\rangle_{FS}\\
    &= \ln\left(\frac{T_{c0}}{T_{c1}}\right)-\left\langle\chi_{B=0}\Phi\left(\rho_{+0},0\right)\right\rangle_{FS}
\end{align}
which leads to a similar form of the gap equation

\begin{equation}
\ln \left(\frac{T}{T_{c0}}\right)+ \left\langle\Phi\left(\rho_{+},Z_{0}\right)+\Phi\left(\rho_{-},Z_{0}\right)-\chi\left[\Phi\left(\rho_{-},Z_{0}\right)-\Phi\left(\rho_{+},Z_{0}\right)\right]\right\rangle_{FS}=0,
\label{general_Bc2_eq_compact}
\end{equation}
and
\begin{equation}
    T_{c1}=T_{c0}e^{-\left\langle\chi_{B=0}\Phi\left(\rho_{+0},0\right)\right\rangle_{FS}}
\end{equation}

\section{Spin-singlet upper critical field $B_{c2}$ -- limiting cases}

\subsection{Useful identities and relations}


In this section we analyze various limits of $B_{c2}$, in particular, the $T\rightarrow 0$ and $T\rightarrow T_{c0}$. We will also consider the small-field limit and the large Ising SOC limit in specific cases. For these purposes, we will find the following identities useful:
\begin{equation}
   \Psi \left( \frac{1}{2} \right)=-2ln2-\gamma, 
\end{equation}
\begin{equation}
   \rm{Re}\left[\Psi\left(z\right)\right]=\frac{1}{2} \left( \Psi\left(z\right)+\Psi\left(z^{*}\right) \right), 
\end{equation}
\begin{equation}
    \lim_{\rho,Z_{0}\to+\infty}\Phi\left(\rho,Z_{0}\right)= \frac{1}{2} \ln \left[ 2e^{\gamma}\sqrt{\left|\rho+2Z_{0}\right|\left|\rho-2Z_{0}\right|} \right],
\end{equation}
\begin{equation}
   \lim_{\rho,Z_{0}\to 0^{+}}\Phi\left(\rho,Z_{0}\right) = -\frac{1}{16}\Psi^{''} \left( \frac{1}{2} \right) \left( \rho^{2}+4Z^{2}_{0} \right) \approx 1.05 \left( \rho^{2}+4Z^{2}_{0} \right). 
\end{equation}
    
When $B\to 0$,
\begin{equation}
   \Phi\left(\rho_-,Z_{0}\right) \rightarrow  1.05 \left( \rho^{2}_-+4Z^{2}_{0} \right) \propto B^2. 
\end{equation}

Additionally, 
\begin{align*}
    \Phi\left(\rho_{+},Z_{0}\right)& \approx \Phi\left(\rho_{+0},0\right)+\frac{i}{8}\left[ \Psi' \left( \frac{1+i\rho_{+0}}{2} \right) - \Psi' \left( \frac{1-i\rho_{+0}}{2} \right) \right] \delta\rho_{+} - \frac{1}{8} \left[ \Psi^{''}\left( \frac{1+i\rho_{+0}}{2} \right) + \Psi^{''} \left( \frac{1-i\rho_{+0}}{2} \right) \right] Z^{2}_{0}\\
    & \equiv  \Phi\left(\rho_{+0},0\right)+f_{1}\left(\rho_{+0}\right)\delta\rho_{+}+f_{2}\left(\rho_{+0}\right)Z^{2}_{0},
\end{align*}
where
\begin{equation}
  \rho_{+0}\equiv\frac{\sqrt{\lambda^{2}_{I}+k^{2}_{F}\lambda^{2}_{R}}}{\pi T_{c}},\quad \delta\rho_+\equiv\rho_+-\rho_{+0},  
\end{equation}

\begin{equation}
     f_{1}\left(\rho_{+0}\right)  \equiv \frac {i}{8} \left[ \Psi' \left( \frac{1+i\rho_{+0}}{2} \right) - \Psi' \left( \frac{1-i\rho_{+0}}{2} \right) \right] =\sum^{\infty}_{n=0}\frac{2\left(2n+1\right)\rho_{+0}}{\left[ \left(2n+1\right)^{2}+\rho^{2}_{+0} \right]^{2}},
\end{equation}
\begin{equation}
    f_{2}\left(\rho_{+0}\right)\equiv-\frac{1}{8} \left[ \Psi^{''} \left( \frac{1+i\rho_{+0}}{2} \right) + \Psi^{''} \left( \frac{1-i\rho_{+0}}{2} \right) \right].
\end{equation}
Both $f_1\left(\rho_{+0}\right)$ and $f_2\left(\rho_{+0}\right)$ are real functions.

\subsection{Limit of $T\rightarrow 0,\; \lambda_I=\lambda_R=g_{\text{orb}}=0$}

In this limit,
\begin{equation}
\begin{gathered}
   P_{\pm}=\mu_B B,\bP_{+}\cdot\bP_{-}=\mu^2_BB^{2},\frac{\bP_{+}\cdot\bP_{-}}{\left|P_{+}\right|\left|P_{-}\right|}=1, \\ 
   Z_{0}=0, \\
   \rho_{+}=\frac{\left|B\right|}{\pi T}\rightarrow +\infty, \\ 
   \Phi\left(\rho_+,0\right)\rightarrow \frac{1}{2}ln \left( 2e^{\gamma}\rho_+ \right), \\ \rho_{-}=0, \\ 
   \Phi \left( \rho_{-},Z_{0} \right)=0. 
\end{gathered}
\end{equation}
Then the equation defining the upper critical field temperature dependence becomes
\begin{equation}
    \ln\left(\frac{T_{c}}{T_{c0}}\right)+2\Phi\left(\rho_{+},0\right)=0.
\end{equation}
At the critical field, $B=\mu_{B}B_{c}$, $T=T_{c}$, and $\rho_{+}=\frac{B_{c}}{\pi T_{c}}.$ Then, the upper critical field at $T=0$ satisfies 
\begin{equation}
\begin{gathered}
   \ln\left(\frac{T_{c}}{T_{c0}}\right)+\ln\left(2e^{\gamma}\rho_{+}\right)=0 \Rightarrow \\ \mu_{B}B_{c}\left(T=0\right)=\frac{\pi}{2}e^{-\gamma}k_BT_{c0} 
   \end{gathered}
\end{equation}

Recall that the Pauli limit is 
\begin{equation}
  B_P=\frac{\Delta_0}{\sqrt{2}\mu_B}=\frac{1.76k_BT_{c0}}{\sqrt{2}\mu_B}.  
\end{equation}
The Pauli-limit violation ratio (PVR) is a constant $B_c/B_P=\pi e^{-\gamma}/\left(1.76\sqrt{2}\right)=1/\sqrt{2}$ (Remind that $1.76=2/1.13=\pi e^{-\gamma}$.). This value is consistent with the numerically obtained result.

\subsection{Large $\lambda_I$ limit with $\lambda_R=g_{\text{orb}}=0$}

In this case, 
\begin{equation}
    \begin{gathered}
        \left|\bP_+\right|^2=\left|\bP_-\right|^2=B^{2}+\lambda^{2}_{I}, \\ \bP_{+}\cdot\bP_{-}=B^{2}-\lambda^{2}_{I}, \\ 
        \chi = \frac{\bP_{+}\cdot\bP_{-}}{\left|P_{+}\right|\left|P_{-}\right|}=\frac{B^{2}-\lambda^{2}_{I}}{B^{2}+\lambda^{2}_{I}},
    \end{gathered}
\end{equation}
therefore,
\begin{equation}
\begin{gathered}
  \rho_{+}=\frac{\left|P_{+}\right|+\left|P_{-}\right|}{2\pi T_{c}}=\frac{\left|P_{+}\right|}{\pi T_{c}},\\  \rho_{-}=0, \\
  \Phi\left(\rho_{+},0\right)=\frac{1}{2}\ln\left(2e^{\gamma}\rho_{+}\right),\\ 
  \Phi\left(\rho_{-}=0,0\right)=0.
  \end{gathered}
\end{equation}
Then, the equation defining the upper critical field temperature dependence becomes
\begin{equation}
  \frac{1+\chi}{2}ln \left( 2e^{\gamma}\rho_{+} \right)=\ln\left(\frac{T_{c0}}{T_{c}}\right).  
\end{equation}

To consider the $\lambda_{I}\gg \mu_BB$ limit we define three parameters:
\begin{equation}
    x\equiv g\mu_BB/\lambda_I,\quad c\equiv k_BT_{c0}/\lambda_I,\quad \delta\equiv T_c/T_{c0}.
\end{equation}
Then
\begin{equation}
   \chi=\frac{x^2-1}{x^2+1},\quad \rho_+=\frac{\sqrt{1+x^2}}{2\pi c\delta}. 
\end{equation}
Exponentiation of both sides yields
\begin{equation}
   \left( \frac{2e^\gamma\sqrt{1+x^2}}{2\pi c\delta} \right)^\frac{x^2}{x^2+1}=\frac{1}{\delta}. 
\end{equation}
Expanding the left-hand-side to order $x^2$, so that $LHS=1-x^2\gamma ln\left(\pi c\delta\right)$, we can solve for $x$:
\begin{equation}
  x=\sqrt{\frac{1-\delta}{\gamma\delta \ln \left( \frac{1}{\pi c\delta} \right)}}.  
\end{equation}
The PVR is then given by 
\begin{equation}
    PVR=\frac{\sqrt{2}}{1.76c}\sqrt{\frac{1-\delta}{\gamma\delta \ln \left( \frac{1}{\pi c\delta} \right) }}=\frac{\sqrt{2}e^\gamma\lambda_I}{\pi gk_B T_{c0}}\sqrt{\frac{T_{c0}-T_c}{\gamma T_c \ln \left( \frac{\lambda_I}{\pi T_c} \right) }}
    \label{PVR_large_Ising_only}
\end{equation}
This approximated formula captures both the square-root feature near the critical temperature and the upturn at low temperatures.

\subsection{Large $\lambda_I$ limit with $T\rightarrow 0, \lambda_{I}\neq 0, \lambda_{R}\neq 0, \lambda_{O}\neq 0$}

To consider this limit, we recall
\begin{equation}
     \Phi\left(\rho_{\pm},Z_{0}\right)\approx\frac{1}{2} \ln \left[ 2e^{\gamma}\sqrt{\left|\rho_{\pm}+2Z_{0}\right|\left|\rho_{\pm}-2Z_{0}\right|} \right].
\end{equation}
In the limit of $T\rightarrow 0$, the equation defining the upper critical field becomes

\begin{equation}
  \ln\left(\frac{T_{c}}{T_{c0}}\right)+ \left\langle \frac{1+\chi}{2}\ln \left[ 2e^{\gamma}\sqrt{\left|\rho_{+}+2Z_{0}\right|\left|\rho_{+}-2Z_{0}\right|} \right]+\frac{1-\chi}{2} \ln \left[ 2e^{\gamma}\sqrt{\left|\rho_{-}+2Z_{0}\right|\left|\rho_{-}-2Z_{0}\right|} \right] \right\rangle_{FS}=0.  
\end{equation}
or
\begin{equation}
\left\langle \frac{1+\chi}{2}\ln \left[ \frac{\sqrt{\left(\left|P_{+}\right|+\left|P_{-}\right|\right)^{2}-4z^{2}_{0}}}{\pi e^{-\gamma}k_BT_{c0}} \right]+\frac{1-\chi}{2} \ln \left[ \frac{\sqrt{\left(\left|P_{+}\right|-\left|P_{-}\right|\right)^{2}-4z^{2}_{0}}}{\pi e^{-\gamma}k_BT_{c0}} \right] \right\rangle_{FS}=0,
\label{general_T=0}
\end{equation}
where
\begin{equation}
\begin{gathered}
   \bP_{+}\cdot\bP_{-}=B^{2}-k^{2}_{F}\lambda^{2}_{R}-\lambda^{2}_{I}, \\ \left|\bP_{\pm}\right|^{2}=\left(B\pm k_{y}\lambda_{R}\right)^{2}+k^{2}_{x}\lambda^{2}_{R}+\lambda^{2}_{I}, \\
   z_{0}=-g_{\text{orb}}\mu_BB\sin\theta,\; \chi \equiv \frac{\bP_{+}\cdot\bP_{-}}{\left|\bP_{+}\right|\left|\bP_{-}\right|}. 
   \end{gathered}
\end{equation}
Here, we introduce two results of the angular average:
\begin{equation}
    \left\langle \ln\left|\sin\theta\right|\right\rangle_{FS}=-\ln2, \quad  \left\langle \ln\sqrt{a^2-b^2\sin^2\theta}\right\rangle_{FS}=\ln\left(\frac{a+\sqrt{a^2-b^2}}{2}\right).
\end{equation}
Performing the same set of steps as in the case of the large $\lambda_I$ limit, we obtain the PVR at $T=0$:
\begin{equation}
    PVR=\frac{\lambda_I}{gB_P}\left(\frac{g^2\lambda_R^2k_F^2}{4g_{\orb}^2\lambda_I^2}-1\right)^{\frac{1}{2}}=\frac{\sqrt{2}e^\gamma\lambda_I}{\pi gk_BT_{c0}}\left(\frac{g^2\lambda_R^2k_F^2}{4g_{\orb}^2\lambda_I^2}-1\right)^{\frac{1}{2}}.
    \label{PVR_large_Ising_T=0_general}
\end{equation}
valid only in the regime of  $g\lambda_Rk_F > 2g_\orb\lambda_I$.

 \subsection{Limiting form near the critical temperature $T\rightarrow T{c0}$ for $\lambda_{I}\neq 0, \lambda_{R}\neq 0, \lambda_{O}\neq 0$}

The linear gap equation quantitatively links the expansion coefficient with the SOC and orbital effect in the clean limit. Therefore, an expansion near the zero-field critical temperature could allow the determination of the SOC and orbital effects through a direct fit.

Near critical temperature $T_{c0}$, the $B_{c2}\left(T\right)$ can be expanded in powers of the in-plane magnetic field $B$ as 
\begin{equation}
T_c \left(B\right) \simeq T_{c0} - c_s B^2,
\end{equation}
or
\begin{equation}
    T_c /T_{c0} \simeq 1 - \tilde{c}_s \text{PVR}^2,
\end{equation}
where $T_{c0}$ is the upper critical temperature of the superconducting phase. For both spin-singlet and spin-triplet superconducting gaps, this behavior matches the result from Ginzburg-Landau theory \cite{Tinkham}. The critical temperature for the spin-singlet case does not change upon adding SOC or orbital effect to the system. The small  $\tilde{c}_s$ indicates the enhancement of the upper critical field $B_{c2}$. We do that by expanding each term in equation \eqref{general_Bc2_eq_compact} at the limit $T_c/T_{c0}\rightarrow 1$ and $B\rightarrow 0$. 

\begin{equation}
\begin{gathered}
   \rho_{+}\rightarrow  \frac{\sqrt{\lambda^{2}_{I}+k^{2}_{F}\lambda^{2}_{R}}}{2\pi T_{c}} + \frac{g^2\left(\lambda^{2}_{I}+k^{2}_{F}\lambda^{2}_{R}\cos^{2}\theta\right)}{4\pi T_{c}\left(\lambda^{2}_{I}+k^{2}_{F}\lambda^{2}_{R}\right)^{3/2}}B^{2}_{c}
     \equiv \rho_{+0}+\delta\rho_{+}, \\
     \rho_{-}\rightarrow\frac{gk_{F}\lambda_{R}\sin\theta}{2\pi T_{c}\sqrt{\lambda^{2}_{I}+k^{2}_{F}\lambda^{2}_{R}}}B_{c},  
\end{gathered}
\end{equation}

\begin{equation}
\begin{gathered}
   \Psi\left(\rho_{+}\right)\rightarrow \Phi\left(\rho_{+0},0\right)+f_{1}\left(\rho_{+0}\right)\delta\rho_{+}+f_{2}\left(\rho_{+0}\right)\frac{g_{\text{orb}}^2B^{2}_{c}}{4\pi^2 T^{2}_{c0}}, \\
   \Psi\left(\rho_{-}\right)\rightarrow 4.207 \left( \frac{\rho^{2}_{-}}{4} + \frac{g_{\text{orb}}^2B^{2}_{c}}{4\pi^2 T^{2}_{c0}} \right) 
\end{gathered}
\end{equation}

\begin{equation}
\begin{gathered}
   \chi\rightarrow -1+\frac{2g^2\left(\lambda^{2}_{I}+k^{2}_{F}\lambda^{2}_{R}\cos^{2}\theta\right)}{\left(\lambda^{2}_{I}+k^{2}_{F}\lambda^{2}_{R}\right)^{2}}B^{2}_{c}, \\
   \ln \left( \frac{T_c}{T_{c0}} \right)\rightarrow -\delta \tilde{T},
   \end{gathered}
\end{equation}

\begin{align*}
    f_{1}\left(\rho_{+0}\right) &\equiv \frac {i}{8} \left[ \Psi' \left( \frac{1+i\rho_{+0}}{2} \right) - \Psi' \left( \frac{1-i\rho_{+0}}{2} \right) \right] =\sum^{\infty}_{n=0}\frac{2\left(2n+1\right)\rho_{+0}}{ \left[ \left(2n+1\right)^{2}+\rho^{2}_{+0} \right]^{2}},\\
    f_{2}\left(\rho_{+0}\right) &\equiv-\frac{1}{8} \left[ \Psi^{''} \left( \frac{1+i\rho_{+0}}{2} \right) + \Psi^{''} \left( \frac{1-i\rho_{+0}}{2} \right) \right].
\end{align*}
Near critical temperature $T_{c1}$, the $H_{c2}\left(T\right)$ can be expanded in powers of the in-plane magnetic field $B$ as 
\begin{equation}
T_c \left(B\right) \simeq T_{c1} - c_s B^2.
\label{Tc_expansion1}
\end{equation}
 
We calculate $c_s$ by expanding each term in Eq. \eqref{general_gap_eq} in small $B\rightarrow 0$ and $\delta\rightarrow0$, where $T_c/T_{c1} = 1 - \delta$ :
\begin{equation}
    \begin{gathered}
        c_{s}=\frac{g^2\left(2\lambda^{2}_{I}+\lambda^{2}_{R}\right)k_B T_{c1}}{\left(\lambda^{2}_{I}+\lambda^{2}_{R}\right)^{2}}\Phi\left(\frac{\sqrt{\lambda^{2}_{I}+k^{2}_{F}\lambda^{2}_{R}}}{2\pi T_{c1}},0\right)\\
    +\frac{0.213}{k_B T_{c1}} \left[ \frac{g^2k^{2}_{F}\lambda^{2}_{R}}{8\left(\lambda^{2}_{I}+k^{2}_{F}\lambda^{2}_{R}\right)}+g^{2}_{\text{orb}} \right].
    \end{gathered}
\end{equation}
A slightly more convenient representation of Eq. \eqref{Tc_expansion1} involves the Pauli limit violation ratio (PVR) and reads  
\begin{equation}
T_c /T_{c1} \simeq 1 - \tilde{c}_s \text{PVR}^2,
\end{equation}
where $\text{PVR} = \frac{B}{B_p}$ with $B_P=1.76k_BT_{c1}/\left(\sqrt{2}\mu_B\right)$ and
\begin{equation}
    \begin{gathered}
        \tilde{c}_s=
        1.56k_B T_{c1}c_{s}
=1.56\frac{g^2\left(2\lambda^{2}_{I}+\lambda^{2}_{R}\right)k_B^2 T^2_{c1}}{\left(\lambda^{2}_{I}+\lambda^{2}_{R}\right)^{2}}\Phi\left(\frac{\sqrt{\lambda^{2}_{I}+k^{2}_{F}\lambda^{2}_{R}}}{2\pi T_{c1}},0\right)
        +0.33 \left[ \frac{g^2k^{2}_{F}\lambda^{2}_{R}}{8\left(\lambda^{2}_{I}+k^{2}_{F}\lambda^{2}_{R}\right)}+g_{\text{orb}}^2 \right].
    \end{gathered}
    \label{cs_PRV}
\end{equation}

\begin{figure}
    \centering
    \includegraphics[width=0.5\linewidth]{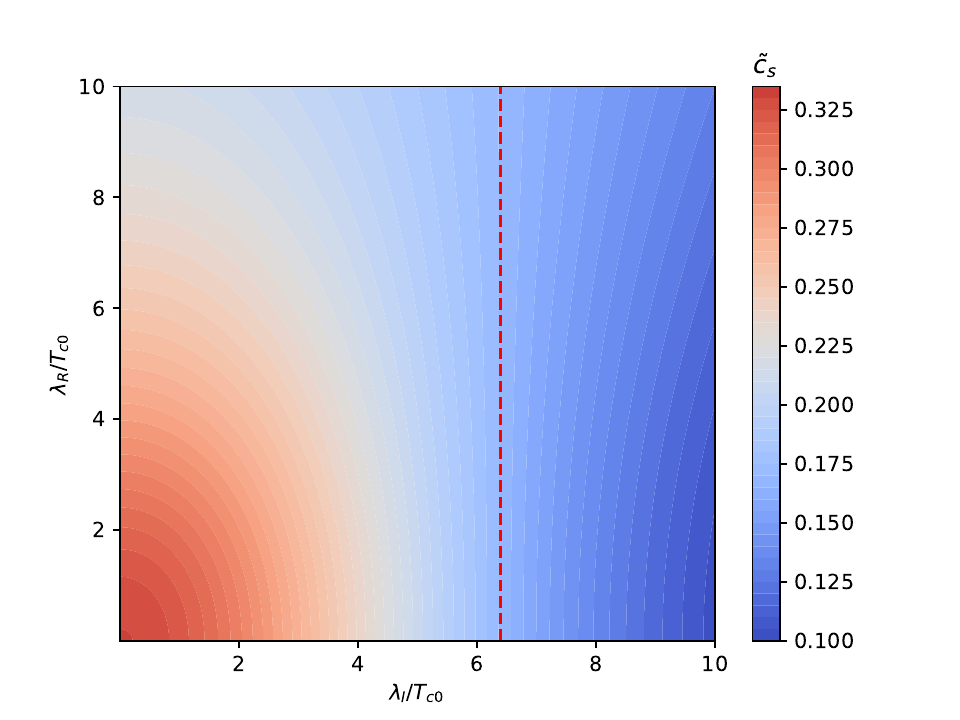}
    \caption{Color map of coefficient $c_s$ without orbital effect. The smaller $\tilde{c}_s$ is, the bigger $B_{c2}$ is expected. When $\lambda_{I}<6.4k_BT_{c0}$, both Ising and Rashba SOC enhance the $B_{c2}$; when $\lambda_{I}>6.4k_BT_{c0}$, Ising SOC enhances the $B_{c2}$, while Rashba SOC against it.}
    \label{fig:color map cs}
\end{figure}

\section{Non-linear correction to identify the phase}

The upper critical curves for spin-triplet pairing presented in the main text are unexpected and far from our impression based on spin-singlet pairing. For that reason, it is helpful to unambiguously distinguish the phase on two sides of the curve, i.e., which region is superconducting and which one is not. Specifically to identify the phase, we need to work out the magnitude of the gap by the non-linear gap equation, but actually we only need to check if the magnitude is 0.\\
Let us recall the derivation of the gap equation. The order parameter can be written as 
\begin{equation}
 \hat{\Delta}= g_{SC}T\sum_{\omega_{n}}\int d^{2}\bs{k}\hat{G}^{0*}_{\omega}\left(-\bs{k}\right)\hat{\Delta}\hat{G}_{\omega}\left(\bs{k}\right),
 \end{equation}
 The Dyson equation shows that
\begin{align*}
    \hat{G}_{\omega}\left(\bs{k}\right) & =\hat{G}^0_{\omega}\left(\bs{k}\right)+\hat{G}^0_{\omega}\left(\bs{k}\right)\hat{\Sigma}\left(\bs{k}\right)\hat{G}^0_{\omega}\left(\bs{k}\right)+\hat{G}^0_{\omega}\left(\bs{k}\right)\hat{\Sigma}\left(\bs{k}\right)\hat{G}^0_{\omega}\left(\bs{k}\right)\hat{\Sigma}\left(\bs{k}\right)\hat{G}^0_{\omega}\left(\bs{k}\right)+\cdots\\
    & =\hat{G}^0_{\omega}\left(\bs{k}\right)\left[I_{2\times 2}-\hat{\Sigma}\left(\bs{k}\right)\hat{G}^0_{\omega}\left(\bs{k}\right)\right]^{-1}
\end{align*}
where $\hat{\Sigma}\left(\bs{k}\right)$ is the self-energy, 
\begin{equation}
    \hat{\Sigma}\left(\bs{k}\right) = \hat{\Delta}^{\dagger}\left(\bs{k}\right)\hat{G}^{0*}_{\omega}\left(-\bs{k}\right)\hat{\Delta}\left(\bs{k}\right).
\end{equation}
Now we can realize that to get the full non-linear equation, we can just replace $ \hat{G}^0_{\omega}\left(\bs{k}\right)$ by $  \hat{G}_{\omega}\left(\bs{k}\right)$. 
\begin{align*}
    \left|\Delta\right|^{2} & = \frac{1}{2} g_{SC}N\left(E_F\right)T\sum_{\omega} \int d\xi \left\langle \text{tr}\left\{\hat{\Delta}^{\dagger}\hat{G}^{0*}_{\omega}\left(-\bs{k}_{F}\right) \hat{\Delta}\hat{G}_{\omega}\left(\bs{k}_{F}\right)\right\}\right\rangle_{FS}\\
    &= \frac{1}{2} g_{SC}N\left(E_F\right)T\sum_{\omega} \int d\xi \left\langle \text{tr}\left\{\hat{\Delta}^{\dagger}\hat{G}^{0*}_{\omega}\left(-\bs{k}_{F}\right) \hat{\Delta}\hat{G}^0_{\omega}\left(\bs{k}_{F}\right)\left[I_{2\times 2}-\hat{\Sigma}\left(\bs{k}_{F}\right)\hat{G}^0_{\omega}\left(\bs{k}_{F}\right)\right]^{-1}\right\}\right\rangle_{FS}\\
    &= \frac{1}{2} g_{SC}N\left(E_F\right)T\sum_{\omega} \int d\xi \left\langle \text{tr}\left\{\hat{\Sigma}\left(\bs{k}_{F}\right)\hat{G}^0_{\omega}\left(\bs{k}_{F}\right)\left[I_{2\times 2}-\hat{\Sigma}\left(\bs{k}_{F}\right)\hat{G}^0_{\omega}\left(\bs{k}_{F}\right)\right]^{-1}\right\}\right\rangle_{FS}\\
    &\approx \frac{1}{2} g_{SC}N\left(E_F\right)T\sum_{\omega} \int d\xi \left\langle \left( \text{tr}\left\{\hat{\Sigma}\left(\bs{k}_{F}\right)\hat{G}^0_{\omega}\left(\bs{k}_{F}\right)\right\}+\text{tr}\left\{\left[\hat{\Sigma}\left(\bs{k}_{F}\right)\hat{G}^0_{\omega}\left(\bs{k}_{F}\right)\right]^{2}\right\}\right)\right\rangle_{FS}\\
\end{align*}
The new term that we didn't calculate is $\text{tr}\left\{\left[\hat{\Sigma}\left(\bs{k}_{F}\right)\hat{G}^0_{\omega}\left(\bs{k}_{F}\right)\right]^{2}\right\}$. To simplify the handwriting, let's take $a_{+}\equiv i\omega-z_{+}=i\omega-\xi+z_{0}$, $a_{-}\equiv -i\omega-z_{-}=-i\omega-\xi-z_{0}$, $b_{+}\equiv \left(i\omega-z_{+}\right)^{2}-\left|\bP_+\right|^2$, $b_{-}\equiv \left(-i\omega-z_{-}\right)^{2}-\left|\bP_-\right|^2$, $\hat{P}_{+}\equiv \bP_{+}\cdot\bs{\sigma}$, $\hat{P}_{-}\equiv \bP_{-}\cdot\bs{\sigma^*}$, $\hat{d}\equiv \bs{d}\cdot\bs{\sigma}$. Then, for the singlet, this trace equals
\begin{equation}
  \text{tr}\left\{\left[\hat{\Sigma}\left(\bs{k}_{F}\right)\hat{G}^0_{\omega}\left(\bs{k}_{F}\right)\right]^{2}\right\}=\frac{2\Delta^{4}}{b^{2}_{+}b^{2}_{-}}\left[a^{2}_{+}a^{2}_{-}+\left|\bP_-\right|^2a^{2}_{+}+\left|\bP_+\right|^2a^{2}_{-}-\left(\bP_{+}\cdot\bP_{-}\right) 4a_{+}a_{-}+\left(2\left(\bP_{+}\cdot\bP_{-}\right)^{2}-\left|\bP_+\right|^2\left|\bP_-\right|^2\right)\right]  
\end{equation}

and for triplet this trace equals to
\begin{equation}
   \text{tr}\left\{\left[\hat{\Sigma}\left(\bs{k}_{F}\right)\hat{G}^0_{\omega}\left(\bs{k}_{F}\right)\right]^{2}\right\}=\frac{\Delta^{4}}{b^{2}_{+}b^{2}_{-}}\text{tr}\left\{ a^{2}_{+}a^{2}_{-}I_{2\times 2}+a^{2}_{+}\hat{P}^{2}_{-}+a^{2}_{-}\hat{P}^{2}_{+}+4a_{+}a_{-}\hat{P}_{+}\hat{d}\sigma_{y}\hat{P}_{-}\sigma_{y}\hat{d}+\left(\hat{P}_{+}\hat{d}\sigma_{y}\hat{P}_{-}\sigma_{y}\hat{d}\right)^{2}\right\} 
\end{equation}

We will now apply it to two cases from the main text.

\subsection{Spin-singlet}

Since neither of the SOC cases we discussed changes the critical temperature for spin-singlet Cooper pairing, we expect the coefficient of the $\Delta^{4}$ term to be positive along the $B=0$ axis. Let us try no SOC and no B to develop intuition with the calculation. In this case, only $a^{2}_{+}a^{2}_{-}$ survives, and
\begin{equation}
    T\sum_{\omega}\int^{+\infty}_{-\infty}d\xi\frac{a^{2}_{+}a^{2}_{-}}{b^{2}_{+}b^{2}_{-}}=T\sum_{\omega}\int^{+\infty}_{-\infty}d\xi\frac{2}{\left(\omega^{2}+\xi^{2}\right)^{2}}=T\sum_{\omega}\sgn\left(\omega\right)\frac{\pi}{\omega^{3}}=\frac{7\zeta\left(3\right)}{4\pi^{2}T^{2}},
\end{equation}
which is positive.

\subsection{Spin-triplet oriented along the $y$-direction, $\lambda_{I}=k_BT_{c0}$}
We now proceed to the unexpected spin-triplet behavior for the spin-triplet pairing discussed in the main text. In this case,
\begin{equation}
    \text{tr}\left[\hat{P}_{+}\hat{d}\sigma_{y}\hat{P}_{-}\sigma_{y}\hat{d}\right]=-2\lambda^{2}_{I}
\end{equation}
\begin{equation}
    \text{tr}\left[\left(\hat{P}_{+}\hat{d}\sigma_{y}\hat{P}_{-}\sigma_{y}\hat{d}\right)^{2}\right]=2\lambda^{4}_{I}
\end{equation}
\begin{equation}
   \int^{+\infty}_{-\infty}d\xi\frac{a^{2}_{+}a^{2}_{-}}{b^{2}_{+}b^{2}_{-}}=\sgn\left(\omega\right)\frac{\pi\left(8\omega^{6}+8\lambda^{2}_{I}\omega^{4}+5\lambda^{4}_{I}\omega^{2}+\lambda^{6}_{I}\right)}{16\omega^{3}\left(\omega^{2}+\lambda^{2}_{I}\right)^{3}} 
\end{equation}
\begin{equation}
    \int^{+\infty}_{-\infty}d\xi\frac{a^{2}_{+}}{b^{2}_{+}b^{2}_{-}}=\sgn\left(\omega\right)\frac{\pi\left(-4\omega^{4}+\lambda^{2}_{I}\omega^{2}+\lambda^{4}_{I}\right)}{16\omega^{3}\left(\omega^{2}+\lambda^{2}_{I}\right)^{3}}
\end{equation}
\begin{equation}
    \int^{+\infty}_{-\infty}d\xi\frac{a^{2}_{-}}{b^{2}_{+}b^{2}_{-}}=\sgn\left(\omega\right)\frac{\pi\left(-4\omega^{4}+\lambda^{2}_{I}\omega^{2}+\lambda^{4}_{I}\right)}{16\omega^{3}\left(\omega^{2}+\lambda^{2}_{I}\right)^{3}}
\end{equation}
\begin{equation}
  \int^{+\infty}_{-\infty}d\xi\frac{4a_{+}a_{-}}{b^{2}_{+}b^{2}_{-}}=\sgn\left(\omega\right)\frac{\pi\left(6\omega^{4}+3\lambda^{2}_{I}\omega^{2}+\lambda^{4}_{I}\right)}{4\omega^{3}\left(\omega^{2}+\lambda^{2}_{I}\right)^{3}}  
\end{equation}
\begin{equation}
   \int^{+\infty}_{-\infty}d\xi\frac{1}{b^{2}_{+}b^{2}_{-}}=\sgn\left(\omega\right)\frac{\pi\left(5\omega^{2}+\lambda^{2}_{I}\right)}{16\omega^{3}\left(\omega^{2}+\lambda^{2}_{I}\right)^{3}} 
\end{equation}

Then the Matsubara sum is
\begin{equation}
 T\sum_{\omega_{n}}\pi\Delta^{4}\sgn\left(\omega_{n}\right)\frac{\omega_{n}\left(\omega_{n}^{2}-3\lambda^{2}_{I}\right)}{\left(\omega_{n}^{2}+\lambda^{2}_{I}\right)^{3}}=2\pi\Delta^{4}T\sum_{\omega_{n}>0}\frac{\omega_{n}\left(\omega_{n}^{2}-3\lambda^{2}_{I}\right)}{\left(\omega_{n}^{2}+\lambda^{2}_{I}\right)^{3}}
\end{equation}

To have a real non-zero solution of $\Delta$, we need the gap equation to have opposite sign of coefficient for $\Delta^{2}$ and $\Delta^{4}$ term.
Recall the result from the previous section, let us write down the equation like
\begin{equation}
    a_{2}\Delta^{2}=a_{4}\Delta^{4}
\end{equation}
where
\begin{equation}
   a_{2}=\ln\left(T/T_{c0}\right)+\left\{ \Phi\left(\rho_{+},Z_{0}\right)+\Phi\left(\rho_{-},Z_{0}\right)-\chi\left[\Phi\left(\rho_{-},Z_{0}\right)-\Phi\left(\rho_{+},Z_{0}\right)\right]\right\} 
\end{equation}
and
\begin{equation}
   a_{4}=\pi T\sum_{\omega_{n}>0}\frac{\omega_{n}\left(\omega_{n}^{2}-3\lambda^{2}_{I}\right)}{\left(\omega_{n}^{2}+\lambda^{2}_{I}\right)^{3}} 
\end{equation}

So if $a_{2}$ and $a_{4}$ have the same sign, then it is in the superconducting phase. 

Now focus on the graphs in the main text. The critical points match with the $B_c$ vs $T_c$ curve. There is a Normal phase at T=0 and B=0, and this is the magnetic field-enhanced SC phase at the intermediate temperature.

\begin{figure}
    \centering
    \includegraphics[width=0.75\linewidth]{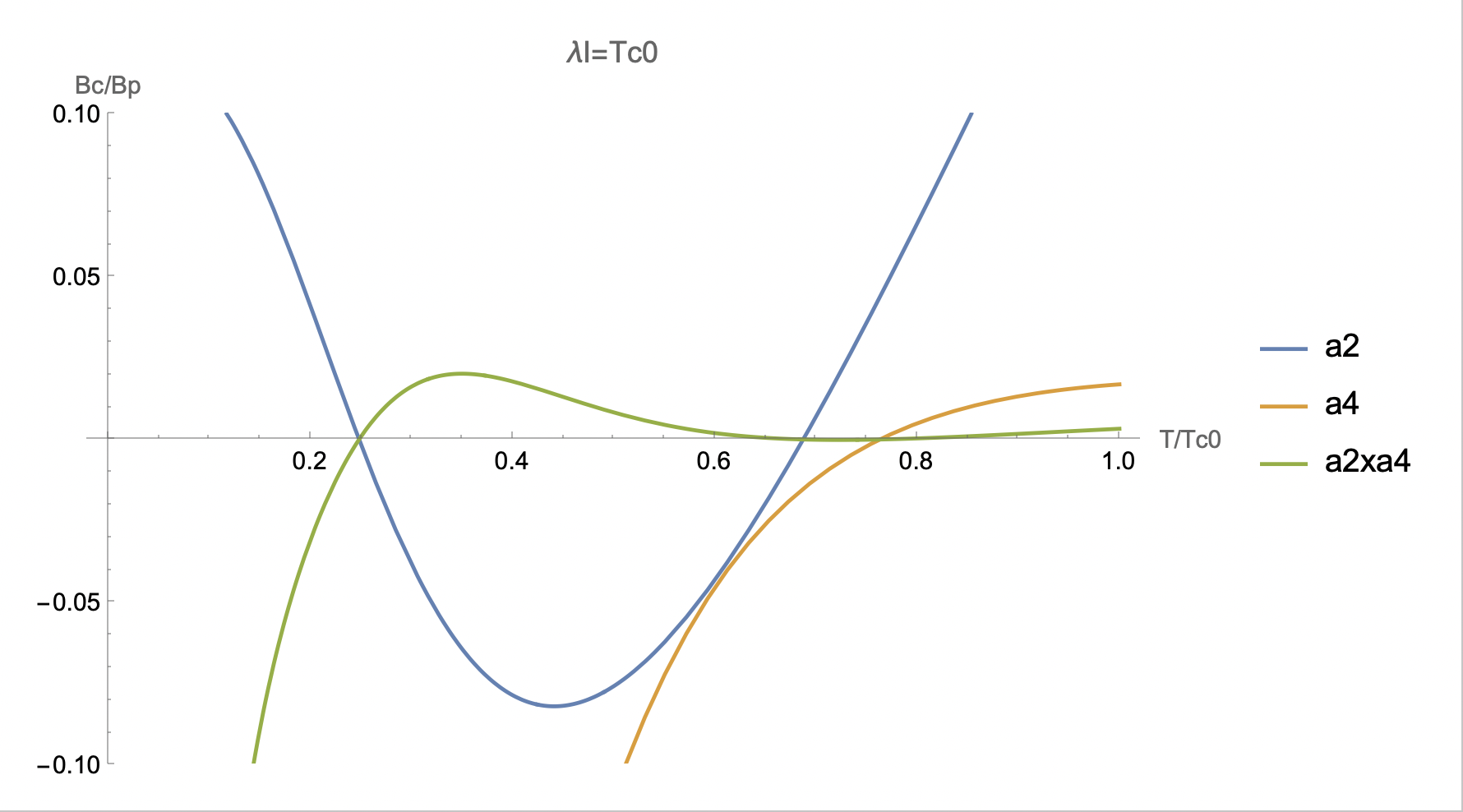}
    \caption{In the near region of the blue curve passing the axis, the green curve above the axis represents a non-zero $\Delta$ and a superconducting phase; otherwise, it represents a normal phase.}
    \label{identify the phase}
\end{figure}

\section{Effective Hamiltonian for Bernal bilayer graphene in sublattice and band basis}

In this section, we relate the microscopic Bernal bilayer graphene (BBG) Hamiltonian to the effective Hamiltonian we are using to calculate the upper critical field. To do so, we employ the Schrieffer–Wolff transformation.  We take the full Hamiltonian of BBG as $H=H^{0}+H^{I}+H^{R}+H^{Z}+H^{O},$ with
\begin{eqnarray}
    H^{0} &=& \begin{bmatrix}
    -U/2 & v\pi^{\dagger} & -v_{4}\pi^{\dagger} & v_{3}\pi \\
    v\pi & -U/2 & \gamma_{1} & -v_{4}\pi^{\dagger} \\
    -v_{4}\pi & \gamma_{1} & U/2 & v\pi^{\dagger} \\
    v_{3}\pi^{\dagger} & -v_{4}\pi & v\pi & U/2
\end{bmatrix} s^{0}, \\
H^{I} &=& \frac{\xi}{2}\begin{pmatrix}
    \lambda_I^{1} & 0 \\
    0 & \lambda_I^{2}
\end{pmatrix}\sigma^{0}s^{z}, \\
H^{R} &=& \frac{1}{2}\begin{pmatrix}
    \lambda_R^{1} & 0 \\
    0 & \lambda_R^{2}
\end{pmatrix}\left(\xi\sigma^{x}s^{y}-\sigma^{y}s^{x}\right), \\
H^{Z} &=& \frac{1}{2}g\mu_B B\sigma^{0}s^{x}, \\ 
H^{O} &=& \alpha \mu_BBl^z\sigma^{y}s^{0}=\frac{edv}{2\mu_B}\mu_B Bl^z\sigma^{y}s^{0},
\end{eqnarray}
where $\left(\xi,l,\sigma,s\right)$ denotes Pauli matrices acting on the valley, layer, sublattice, and spin degrees of freedom respectively,  $l$ labels layers, $\pi=\xi p_{x}+ip_{y}$, and  $\xi$ is the valley index. We consider here that only one layer (top) has proximitized SOC, whereas the other (bottom) layer experiences no substrate-induced SOC.

Next, we find the form of these effective spin Hamiltonians and orbital effects in the low-energy space. In our perturbation analysis, we take the limit $v\left|\bs{p}\right|\ll U\ll \gamma_1,\; v_3\ll v.$ First, when we project the Hamiltonians above into A1B2 sublattice space (corresponding to the effective basis of the large $U$ system), to yield: 
\begin{align}
    H^{0}_{eff}=\begin{bmatrix}
    -\frac{U}{2} +\frac{v^{2}p^{2}U}{2\gamma^{2}_{1}} & -\frac{v^{2}\pi^{\dagger 2}}{\gamma_{1}}+v_{3}\pi  \\
    -\frac{v^{2}\pi^{2}}{\gamma_{1}}+v_{3}\pi^{\dagger} & \frac{U}{2} -\frac{v^{2}p^{2}U}{2\gamma^{2}_{1}} \\
\end{bmatrix}
s^{0},
\end{align}

\begin{equation}
    H^{I}_{eff}=\frac{\xi}{2}\lambda_{I}\begin{bmatrix}
    1 & 0  \\
    0 & \frac{v^{2}p^{2}}{\gamma^{2}_{1}}
\end{bmatrix}
s^{z},
\end{equation}

\begin{equation}
    H^{R}_{\eff}=\frac{1}{2}g_{R}\begin{bmatrix}
    0 & 0 & 0 & \frac{i\left(1-\xi\right)v\left(p_x+ip_y\right)}{\gamma_{1}} \\
    0 & 0 & \frac{-i\left(1+\xi\right)v\left(p_x-ip_y\right)}{\gamma_{1}} & 0 \\
    0 & \frac{i\left(1+\xi\right)v\left(p_x+ip_y\right)}{\gamma_{1}} & 0 & 0 \\
    \frac{-i\left(1-\xi\right)v\left(p_x-ip_y\right)}{\gamma_{1}} & 0 & 0 & 0
\end{bmatrix}
\end{equation}
\begin{equation}
    H^{Z}_{\eff}=\frac{1}{2}g\mu_B B\left(1+\frac{v^{2}p^{2}}{\gamma^{2}_{1}}\right)\sigma^0 s^{x},
\end{equation}
\begin{equation}
H^{O}_{\eff}=\frac{\alpha Uvp_{y}}{\gamma^{2}_{1}}\mu_B B\sigma_0 s^{0}.
\end{equation}

Next, we project to the band basis. After we derive the full Hamiltonian by Schrieffer–Wolff transformation, the projection to the band basis here is just a unitary transformation in the sublattice space. We need to perform the same unitary transformation to diagonalize $H^0_{\eff}$. Notice that $H^0_{\eff}$ is in the form
\begin{equation}
  H^0_{\eff}=\bs{m}\cdot\bs{\sigma}\otimes s^0,  
\end{equation}
so the unitary matrix has the form
\begin{equation}
U=\begin{bmatrix}
    \cos\left(\theta/2\right) & -e^{-i\phi}\sin\left(\theta/2\right)\\
    e^{i\phi}\sin\left(\theta/2\right) & \cos\left(\theta/2\right)
\end{bmatrix},
\end{equation}
where $\bs{m}=\left|\bs{m}\right|\left(\sin\theta \cos\phi,\;\sin\theta \sin\phi,\;\cos\theta\right)$, and
\begin{equation}
   m_x= v^2\left(-p_x^2 + p_y^2\right)/\gamma_1 + \xi v_3p_x,\quad m_y=-2\xi v^2p_xp_y/\gamma_1 - v_3p_y,\quad m_z=\left[v^2\left(p_x^2 + p_y^2\right)/\gamma_1^2 - 1\right]U/2. 
\end{equation}
\begin{equation}
   \sin\theta=\frac{\sqrt{m^2_x+m^2_y}}{\left|\bs{m}\right|},\quad \cos\theta=\frac{m_z}{\left|\bs{m}\right|},\quad \sin\phi=\frac{m_y}{\sqrt{m^2_x+m^2_y}},\quad \cos\phi=\frac{m_x}{\sqrt{m^2_x+m^2_y}}.
\end{equation}
After the transformation $H_{\band}=U^\dagger H_{eff} U$, all of the elements only include $\sin\theta$ and $\cos\theta$, not $\sin\left(\theta/2\right)$ and $\cos\left(\theta/2\right)$. We can restore the trigonometric function values at the end of the calculation.

Finally, we project to the basis $\left(c\uparrow,c\downarrow,v\uparrow,v\downarrow\right)$, where c and v mean conducting and valence band. We find that the leading SOC correction is diagonal in the band basis, with the Zeeman and orbital effects unchanged.
\begin{equation}
    H^{Z}_{\band}=\left(1+\frac{v^2p^2}{\gamma^2_1}\right)\mu_B B\sigma_0  s^x,    
\end{equation}
\begin{equation}
    H^{O}_{\band}=\frac{Uedv^2p_y}{2\mu_B\gamma^2_1}\mu_BB\sigma_0 s^0=g_\orb\mu_BB\sigma_0 s^0.    
\end{equation}
We stress that the proximity nature of Ising and Rashba SOC makes both projected effective Hamiltonians band-dependent and displacement-field-dependent. 

The Fermiology study of the parent state of the SC2 pocket, e.g., Ref. ~\cite{Holleis2025_Nemacity_BBG}, suggests that the electrons near the Fermi surface localize at 3 trigonal warping pockets and undergo a nematic phase transition. Therefore, we choose to project the effective Hamiltonian to the band basis near one of the trigonal warping centers. Here, we chose a trigonal warping pocket along the $x$ axis; results for other pockets can be derived accordingly. Defining $\bs{p}\equiv \bs{k}-\bs{k}_{0}$ as the momentum relative to the trigonal wrapping minimum, where $\bs{k}_{0}$ is the vector from the K(K') point to the trigonal wrapping minimum.
\begin{equation}
    k_0=\frac{3\gamma_1^3v_3 +\gamma_1\sqrt{
   U^2\left(4\gamma_1^2+U^2\right)v^2+\gamma_1^2 \left(\gamma_1^2-2U^2\right) v_3^2}}{\left(4\gamma_1^2 + 
   U^2\right)v^2}\approx \frac{3\gamma_1v_3+\sqrt{4 U^2 v^2+\gamma_1^2v_3^2}}{4v^2}
\end{equation}
In the trigonal warping pocket along the $k_x-$ axis, the orbital effect does not enter as $k_{0y}=0$ - this is a consequence of the in-plane magnetic field being directed along the same axis. The Zeeman term has the same form as above, but with the characteristic momentum set by $k_0$,
\begin{equation}
   H^{Z}_{\band}=\left(1+\frac{v^2k^2_0}{\gamma^2_1}\right)\mu_B B\sigma_0 s_x 
\end{equation}
Here $vk_0/\gamma_1$ is in the same order of $v_3/v$, so the renormalization of the g factor is no bigger than 8\%  as discussed in the main text. 

The Ising term takes the form:
\begin{equation}
    H^{I}_{band}=\frac{\xi\lambda_I}{2}\begin{bmatrix}
   \frac{1-\sgn\left(U\right)}{2} & 0 & f\left(U\right) & 0  \\
    0 & -\frac{1-\sgn\left(U\right)}{2} & 0 & -f\left(U\right)\\
     f^*\left(U\right) & 0 & \frac{1+\sgn\left(U\right)}{2} & 0  \\
    0 & -f^*\left(U\right) & 0 & -\frac{1+\sgn\left(U\right)}{2}
\end{bmatrix},
\end{equation}
where 
\begin{equation}
    f\left(U\right)=\frac{2k_0\left(\gamma_1v_3-4ik_yv^2\right)+U^2}{4\gamma_1\left|U\right|}+\frac{\xi k_x}{\gamma_1\left|U\right|}\left(2v^2k_0-\gamma_1v_3-2ik_yv^2\right)
\end{equation}
Note that the bare Ising value within each band remains unchanged by the effective projection to the low-energy subspace.

For the Rashba Hamiltonian projected on two bands, the leading term turns out to be
\begin{align}
   H^{R}_{conduction}=\frac{1}{2}\lambda_R\left(g_{R,y}s_y-g_{R,x}s_x\right),\\
   H^{R}_{valence}=\frac{1}{2}\lambda_R\left(g_{R,y}s_y+g_{R,x}s_x\right),\\ 
\end{align}
where 
\begin{align}
\label{effective_rashba}
g_{R,x}\left(\bs{k}\right) &\approx \lambda_R^{l=1} \frac{k_y v \left(14 \gamma_1 k_0 v_3+U^2\right)}{2 \gamma_1^2 \left|U\right|},\\
g_{R,y}\left(\bs{k}\right) &\approx \lambda_R^{l=1} \left[ \frac{\xi k_0 \left|U\right| v}{2 \gamma_1^2} +\frac{k_x  v \left(10 \gamma_1 k_0 v_3+3 U^2\right)}{2 \gamma_1^2 \left|U\right|}\right].
\end{align}
The magnitude of Rashba SOC does not show band dependence and is renormalized by $ \frac{ k_0 \left|U\right| v}{2 \gamma_1^2} $, which makes it less relevant at large $U$ than the Ising term.

\section{Fitting to the experimental data}

\begin{figure}[h!]
    \centering
    \includegraphics[width=\linewidth]{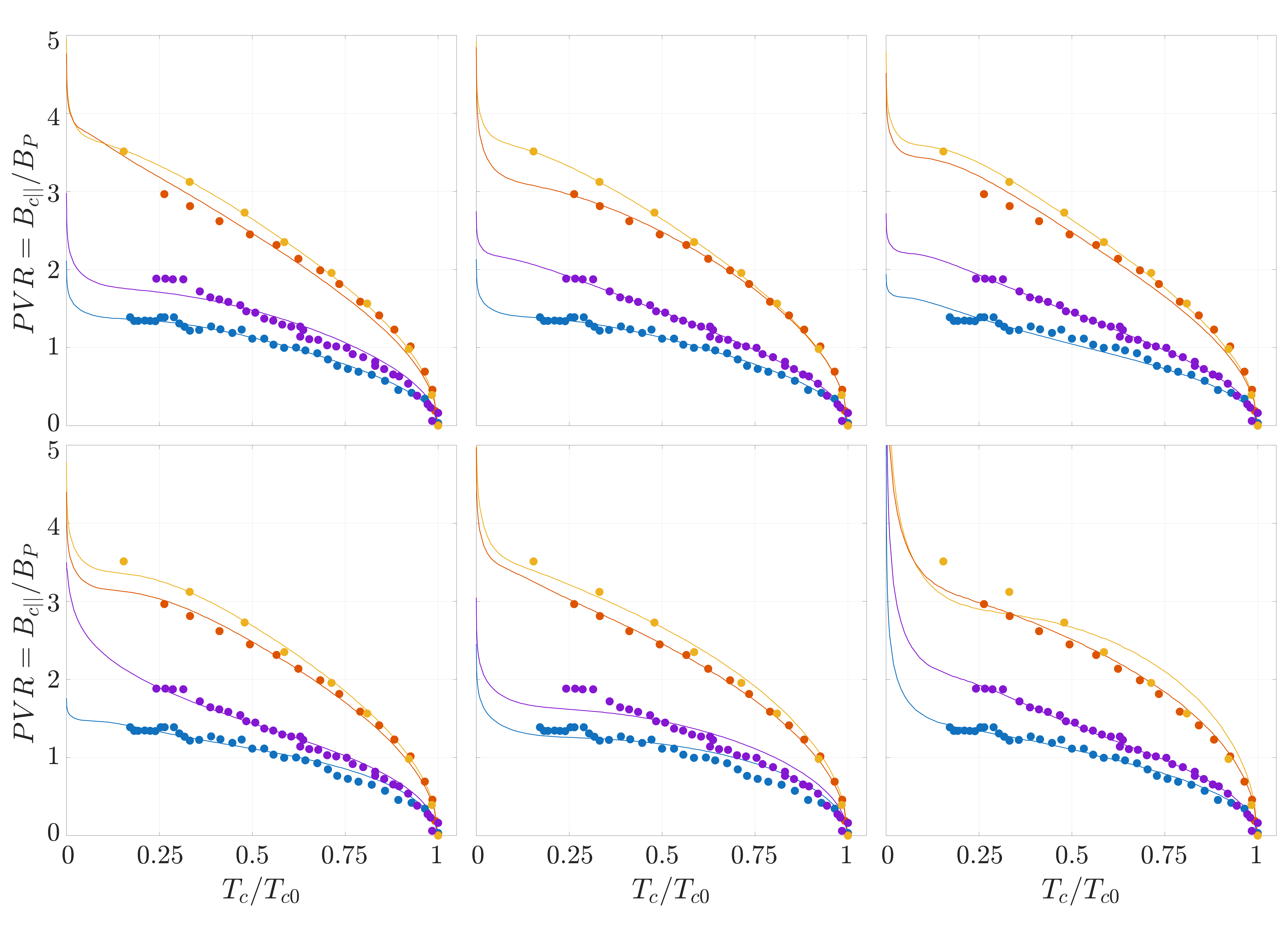}
    \caption{The Pauli-limit violation ratio (PVR) fitting to the experimental data from \textbf{a}\cite{Zhang2023_Enhanced_SC_proximitized_BBG}, \textbf{b}\cite{Zhang2025-ty}, \textbf{c}\cite{Holleis2025_Nemacity_BBG}, and \textbf{d}\cite{Tingxin_Li2024_BBG_WSe2}}
    \label{fig:curve fitting}
\end{figure}

We take the upper critical fields data from the Ref. \cite{Zhang2023_Enhanced_SC_proximitized_BBG,Zhang2025-ty,Holleis2025_Nemacity_BBG,Tingxin_Li2024_BBG_WSe2}, which are believed to be inside the SC2 pocket. We test different theoretical models for the dependence of the effective SOC parameters by fitting.  In model 1, we allow the fit to vary the Rashba SOC $\lambda_R$, inter-layer orbital effect $g_{\text{orb}}$, and  Landé g factor (with $\lambda_I$ set by experimental fit to Landau level crossing). In model 2, we allow the fit to vary the Ising SOC. In model 3, we assume $\lambda_R=0$ and allow the other parameters to vary. In models 4, 5, and 6, we allow the $g$ factor to vary and use the experimentally expected $\lambda_I$. The summary is shown in \ref{tab: model 1 fitting result}, and curve fitting is shown in \ref{fig:curve fitting}.  Here, we show that the effective Rashba effect is irrelevant to the electron pocket in the BBG, consistent with the result of the effective model derived from microscopic projection. The $g$ factors, on the other hand, are unexpectedly large in the fit. This result may be due to the interaction enhancement of the Zeeman energy scale \cite{LarentisMoSe2} or the mismatch of the BKT transition temperature $T_c^{BKT}$ with the BCS critical temperature $T_c^{BCS}$. The overestimated critical temperature will lead to smaller fitted SOC values than expected, as we saw in model 2. Finally, the models that assume $g=2$ are worse than the other fits, pointing to the interaction-enhanced $ g$-factor as a possible explanation. 

\begin{table}[h!]
    \centering
    \begin{tabular}{c|cccccccc}
         &  n (10$^{11}$ cm$^{-2}$)&  D (V/nm)&  U (eV) &$T_{c0}$ (K)&  $\lambda_I$ (meV)&  $\lambda_R$ (meV)&  $g_{\text{orb}}$& $g/g_0$\\
         \hline
         \textbf{a}&  -7&  1.1&  0.11 &0.280&  0.7&  0.0893&  0.0273& 4.3568\\
         \textbf{b}&  -7.3&  1.2&  0.12 &0.391&  1.5&  0.0461&  0.1225& 3.1193\\
         \textbf{c}&  -7.3&  1.15&  0.115 &0.277&  1.6&  0.0301&  0.1470& 4.0444\\
         \textbf{d}&  -5.9&  0.96&  0.096 &0.193&  1.7&  0.1343&  0.1560& 6.8332\\
    \end{tabular}
    \caption{Fitting results of the upper critical field in SC2 pockets in BBG/WSe$_2$. The data are from \textbf{a}\cite{Zhang2023_Enhanced_SC_proximitized_BBG}, \textbf{b}\cite{Zhang2025-ty}, \textbf{c}\cite{Holleis2025_Nemacity_BBG}, and \textbf{d}\cite{Tingxin_Li2024_BBG_WSe2}, the fitting is done using the model 1 approach by using the expected $\lambda_I$ values.}
    \label{tab: model 1 fitting result}
\end{table}
  
\begin{table}[h!]
    \centering
    \begin{tabular}{c|cccccccc}
         &  n (10$^{11}$ cm$^{-2}$)&  D (V/nm)&  U (eV) &$T_{c0}$ (K)&  $\lambda_I$ (meV)&  $\lambda_R$ (meV)&  $g_{\text{orb}}$& $g/g_0$\\
         \hline
         \textbf{a}&  -7&  1.1&  0.11 &0.280&  0.5293&  0.0874&  0.1340& 3.4565\\
         \textbf{b}&  -7.3&  1.2&  0.12 &0.391&  1.3778&  0.1495&  0.0943& 2.5193\\
         \textbf{c}&  -7.3&  1.15&  0.115 &0.277&  1.0543&  0.0280&  0.1488& 2.8262\\
         \textbf{d}&  -5.9&  0.96&  0.096 &0.193&  0.3843&  0.0132&  0.1638& 2.5609\\
    \end{tabular}
    \caption{Model 2 fitting result allowing the fit to vary all parameters of the effective model.}
    \label{tab: model 2 fitting result}
\end{table}

\begin{table}[h!]
    \centering
    \begin{tabular}{c|cccccccc}
         &  n (10$^{11}$ cm$^{-2}$)&  D (V/nm)&  U (eV) &$T_{c0}$ (K)&  $\lambda_I$ (meV)&  $\lambda_R$ (meV)&  $g_{\text{orb}}$& $g/g_0$\\
         \hline
         \textbf{a}&  -7&  1.1&  0.11 &0.280&  0.1310&  0&  0.1144& 1.5178\\
         \textbf{b}&  -7.3&  1.2&  0.12 &0.391&  0.5647&  0&  0.1142& 1.4171\\
         \textbf{c}&  -7.3&  1.15&  0.115 &0.277&  0.4509&  0&  0.1186& 1.4549\\
         \textbf{d}&  -5.9&  0.96&  0.096 &0.193&  0.2237&  0&  0.1181& 1.7211\\
    \end{tabular}
    \caption{Model 3 fitting result, here we set $\lambda_R=0$ and allow all other fit parameters to vary.}
    \label{tab: model 3 fitting result}
\end{table}

\begin{table}[h!]
    \centering
    \begin{tabular}{c|cccccccc}
         &  n (10$^{11}$ cm$^{-2}$)&  D (V/nm)&  U (eV) &$T_{c0}$ (K)&  $\lambda_I$ (meV)&  $\lambda_R$ (meV)&  $g_{\text{orb}}$& $g/g_0$\\
         \hline
         \textbf{a}&  -7&  1.1&  0.11 &0.280&  0.0663&  0&  0.1795& 1\\
         \textbf{b}&  -7.3&  1.2&  0.12 &0.391&  0.3844&  0&  0.1719& 1\\
         \textbf{c}&  -7.3&  1.15&  0.115 &0.277&  0.3054&  0&  0.1672& 1\\
         \textbf{d}&  -5.9&  0.96&  0.096 &0.193&  0.0977&  0&  0.0011& 1\\
    \end{tabular}
    \caption{Model 4 fitting result, here we set $\lambda_R=0$ and $g=2$.}
    \label{tab: model 4 fitting result}
\end{table}

\begin{table}[h!]
    \centering
    \begin{tabular}{c|cccccccc}
         &  n $(10^{11} cm^{-2}$)&  D (V/nm)&  U (eV) &$T_{c0}$ (K)&  $\lambda_I$ (meV)&  $\lambda_R$ (meV)&  $g_{\text{orb}}$& $g/g_0$\\
         \hline
         \textbf{a}&  -7&  1.1&  0.11 &0.280&  0.0393&  0.1557&  0.1324& 1\\
         \textbf{b}&  -7.3&  1.2&  0.12 &0.391&  0.4198&  0.1188&  0.1439& 1\\
         \textbf{c}&  -7.3&  1.15&  0.115 &0.277&  0.3713&  0.1207&  0.1497& 1\\
         \textbf{d}&  -5.9&  0.96&  0.096 &0.193&  0.1287&  0.1303&  0.1633& 1\\
    \end{tabular}
    \caption{Model 5 fitting result, here we set $g=2$.}
    \label{tab: model 5 fitting result}
\end{table}

\begin{table}[h!]
    \centering
    \begin{tabular}{c|cccccccc}
         &  n $(10^{11} cm^{-2}$)&  D (V/nm)&  U (eV) &$T_{c0}$ (K)&  $\lambda_I$ (meV)&  $\lambda_R$ (meV)&  $g_{\text{orb}}$& $g/g_0$\\
         \hline
         \textbf{a}&  -7&  1.1&  0.11 &0.280&  0.7&  0.7179&  1.2968& 1\\
         \textbf{b}&  -7.3&  1.2&  0.12 &0.391&  1.5&  0.4341&  0.5269& 1\\
         \textbf{c}&  -7.3&  1.15&  0.115 &0.277&  1.6&  0.1917&  0.4936& 1\\
         \textbf{d}&  -5.9&  0.96&  0.096 &0.193&  1.7&  1.6871&  1.0382& 1\\
    \end{tabular}
    \caption{Model 6 fitting result, here we set $g=2$ and use the experimentally expected $\lambda_I$.}
    \label{tab: model 6 fitting result}
\end{table}

\end{document}